\documentclass[prd,superscriptaddress,amsfonts,amssymb,amsmath,showpacs,twocolumn]{revtex4-2}

\usepackage{bm}
\usepackage{amsfonts}
\usepackage{latexsym}
\usepackage[latin1]{inputenc}
\usepackage{graphicx}
\usepackage{amsmath}
\usepackage{palatino}
\usepackage{mathpazo}
\usepackage{textcomp}
\usepackage{float}
\usepackage{booktabs}
\usepackage{dcolumn}
\usepackage{ragged2e}
\usepackage{hyperref}
\hypersetup{colorlinks,citecolor=blue}
\hypersetup{colorlinks=true,linkcolor=red,filecolor=magenta,    urlcolor=blue}
\usepackage{amsmath}
\usepackage{graphicx}% Include figure files
\usepackage{makecell}
\usepackage{hyperref}% add hypertext capabilities
\usepackage[mathlines]{lineno}% Enable numbering of text and display math
\usepackage{amssymb}
\usepackage{enumitem}
\usepackage{mathtools}
\usepackage{mathrsfs}
\usepackage{setspace}
\usepackage{color}
\usepackage{graphicx}
\usepackage[dvipsnames]{xcolor}
\usepackage{mathalfa}
\usepackage{calligra}
\usepackage[misc]{ifsym}
\DeclareMathAlphabet{\mathpzc}{OT1}{pzc}{m}{it}
\DeclareMathAlphabet{\mathcalligra}{T1}{calligra}{m}{n}
\hypersetup{colorlinks=true,citecolor=blue,linkcolor=red,urlcolor=magenta}
\usepackage{xcolor}
\usepackage{orcidlink}
\usepackage[caption=false]{subfig}
\usepackage{commath}
\def\jnl@style{}
\def\aaref@jnl#1{{\jnl@style#1}}

\def\aaref@jnl#1{{\jnl@style#1}}

\def\aj{\aaref@jnl{AJ}}                   % Astronomical Journal
\def\apj{\aaref@jnl{ApJ}}                 % Astrophysical Journal
\def\apjl{\aaref@jnl{ApJ}}                % Astrophysical Journal, Letters
\def\apjs{\aaref@jnl{ApJS}}               % Astrophysical Journal, Supplement
\def\apss{\aaref@jnl{Ap\&SS}}             % Astrophysics and Space Science
\def\aap{\aaref@jnl{A\&A}}                % Astronomy and Astrophysics
\def\aapr{\aaref@jnl{A\&A~Rev.}}          % Astronomy and Astrophysics Reviews
\def\aaps{\aaref@jnl{A\&AS}}              % Astronomy and Astrophysics, Supplement
\def\mnras{\aaref@jnl{Mon.~Not.~Roy.~Astron.~Soc.}}             % Monthly Notices of the RAS
\def\prd{\aaref@jnl{Phys.~Rev.~D}}        % Physical Review D
\def\plb{\aaref@jnl{Phys.~Lett.~B}}        % Physics Letters B
\def\prc{\aaref@jnl{Phys.~Rev.~C}}  % Physical Review C
\def\prl{\aaref@jnl{Phys.~Rev.~Lett.}}    % Physical Review Letters
\def\qjras{\aaref@jnl{QJRAS}}             % Quarterly Journal of the RAS
\def\skytel{\aaref@jnl{S\&T}}             % Sky and Telescope
\def\ssr{\aaref@jnl{Space~Sci.~Rev.}}     % Space Science Reviews
\def\zap{\aaref@jnl{ZAp}}                 % Zeitschrift fuer Astrophysik
\def\nat{\aaref@jnl{Nature}}              % Nature
\def\aplett{\aaref@jnl{Astrophys.~Lett.}} % Astrophysics Letters
\def\apspr{\aaref@jnl{Astrophys.~Space~Phys.~Res.}} % Astrophysics Space Physics Research
\def\physrep{\aaref@jnl{Phys.~Rep.}}      % Physics Reports
\def\physscr{\aaref@jnl{Phys.~Scr}}       % Physica Scripta
\def\commat{\aaref@jnl{Comm.~Math.~Phys.}}              % Communications in Mathematical Physics
\def\science{\aaref@jnl{Science}}               % Science
\def\cqg{\aaref@jnl{Classical Quant.~Grav.}}            % Classical and Quantum Gravity
\def\jpcs{\aaref@jnl{JPCS}}                                     % Journal of Physics Conference Series
\def\ijmpd{\aaref@jnl{Int.~J.~Mod.~Phys.~D}}                    % International Journal of Modern Physics D
\def\grg{\aaref@jnl{Gen.~Relat.~Gravit.}}               % General Relativity and Gravitation
\def\rpp{\aaref@jnl{Rep.~Prog.~Phys.}}          % Reports on Progress in Physics
\def\npa{\aaref@jnl{Nucl.~Phys.~A}}        % Nuclear Physics A
\def\lrr{\aaref@jnl{Living Rev.~Rel.}}                   % Living reviews in relativity
\def\jcap{\aaref@jnl{J.~Cosmology Astropart.~Phys.}}    % Journal of cosmology and astroparticle physics
\def\rmp{\aaref@jnl{Rev.~Mod.~Phys.}}   %Reviews of modern physics
\def\epjc{\aaref@jnl{Eur.~Phys.~J.~C}}

\allowdisplaybreaks[1]
\renewcommand{\arraystretch}{1.1}
\begin{document}

\preprint{APS/123-QED}

\title{Static traversable wormhole solutions in $\mathpzc{f}(\mathcal{R},\mathscr{L}_m)$ gravity}% Force line breaks with \\


\author{N. S. Kavya\orcidlink{0000-0001-8561-130X}}
\email{kavya.samak.10@gmail.com}
\affiliation{Department of P.G. Studies and Research in Mathematics,
 \\
 Kuvempu University, Shankaraghatta, Shivamogga 577451, Karnataka, INDIA
}%

\author{V. Venkatesha\orcidlink{0000-0002-2799-2535}}%
 \email{vensmath@gmail.com}
\affiliation{Department of P.G. Studies and Research in Mathematics,
 \\
 Kuvempu University, Shankaraghatta, Shivamogga 577451, Karnataka, INDIA
}%

\author{G. Mustafa\orcidlink{0000-0003-1409-2009}}%
\email{gmustafa3828@gmail.com}
\affiliation{
 Department of Physics, Zhejiang Normal University, Jinhua, 321004, People's Republic of China.
}%

\author{P.K. Sahoo\orcidlink{0000-0003-2130-8832}}
\email{pksahoo@hyderabad.bits-pilani.ac.in}
\affiliation{
 Department of Mathematics, Birla Institute of Technology and Science-Pilani,\\
 Hyderabad Campus, Hyderabad 500078, INDIA
}%

\author{S. V. Divya Rashmi}
\email{rashmi.divya@gmail.com}
\affiliation{Department of Mathematics, Vidyavardhaka College of Engineering,\\
Mysuru - 570002, INDIA}


\date{\today}% It is always \today, today,

\begin{abstract}
In this study, we explore the new wormhole solutions in the framework of new modified $\mathpzc{f}(\mathcal{R},\mathscr{L}_m)$ gravity. To obtain a characteristic wormhole solution, we use anisotropic matter distribution and a specific form of energy density. As second adopt the isotropic case with a linear EoS relation as a general technique for the system and discuss several physical attributes of the system under the wormhole geometry. Detailed analytical and graphical discussion about the matter contents via energy conditions is discussed. In both cases, the shape function of wormhole geometry satisfies the required conditions. Several interesting points have evolved from the entire investigation along with the features of the exotic matter within the wormhole geometry. Finally, we have concluding remarks. 
\begin{description}
\item[Keywords]
Wormhole, $\mathpzc{f}(\mathcal{R},\mathscr{L}_m)$ gravity, energy conditions.

\end{description}
\end{abstract}

\maketitle

%\tableofcontents
\section{INTRODUCTION}\label{sectionI}
	\par Over the past two decades, probing wormhole solutions has become a focus of interest in contemporary astronomy. In 1916, Flamm, as a pioneer, provided a mathematical notion of this hypothetical structure \cite{flamm}. To overcome the instability in this solution, Einstein and Rosen proposed a bridge-like structure known as the Einstein-Rosen bridge \cite{erbridge}. Later, the humanly traversable wormhole was witnessed by the mathematically derived intuition of Morris and Thorne \cite{morrisandthorne} in which the structure ruled out the presence of the event horizon. But, these wormholes violated null energy conditions (NEC). For an ordinary matter, a violation of NEC breaks the laws of physics. With this motivation, the existence of a hypothetical fluid known as the `exotic matter' was taken into account and hence the traversability was achieved. However, the physical reality of such wormholes was put unanswered. Then, numerous initiatives were made to address this issue \cite{em1}. 
	
	\par One of the promising ways to deal with the exotic fluid problem is to use the modified theoretic approach. This can reduce or even can nullify the usage of exotic matter \cite{ec1,ec2,ec3,ec4,ec5,ec6}. For instance, in \cite{ec2} authors have constructed the wormhole solutions in the context of $\mathpzc{f}(\mathcal{R})$ gravity that satisfies the NEC. In \cite{ref1}, Maldacena and Milekhin explored humanly traversable wormholes with the Randall-Sundram model. Numerous endeavors have been taken to examine the nature of wormholes in higher dimensions \cite{ref2,ref3,ref4,ovgun}. Rahaman et al. investigated the possible existence of wormholes in the galactic halo region \cite{rahaman}. Dynamics of spherically symmetric traversable wormholes possessing thin shells are investigated in \cite{ref5}.  As a quantum field approach traversable wormholes are studied in four-dimension, whose solution can be embedded in the standard wormhole with appropriate conditions \cite{ref6}. Ref. \cite{ref7} gives the studies on the stable wormhole in exponential $\mathpzc{f}(\mathcal{R})$ gravity. Several interesting works on wormhole solutions within the background of different modified theories can be seen in the literature. Say, in $\mathpzc{f}(\mathcal{R})$ \cite{fr}, Gauss-Bonnet \cite{gb}, brane \cite{brane}, teleparallel \cite{t}, symmetric teleparallel \cite{st}, modified teleparallel \cite{mt}, $\mathpzc{f}(\mathcal{R},\mathcal{T})$ \cite{rt}, Rastall gravity \cite{rastall}. Moreover, in evaluating the universe, $\mathpzc{f}(\mathcal{R})$ gravity produced a reliable framework. It can adequately explain the late-time acceleration \cite{f1}, the flatness of rotational curves of the galaxies \cite{f2}, the unification of inflation with dark energy \cite{f3} and the analysis of galactic dynamics of massive test particles without dark matter \cite{f4}. With this motivation, many generalized $\mathpzc{f}(\mathcal{R})$ theories were proposed.
	
    \par Among the generalized $\mathpzc{f}(\mathcal{R})$ theories, the theory that has an explicit matter coupling with the curvature is  $\mathpzc{f}(\mathcal{R},\mathscr{L}_m)$ gravity \cite{frlm}. The primary merit of this gravity is the generalization of both geometry and matter elements of the theory. Due to the coupling, the test particles obey the non-geodesic motion which results in the violation of the equivalence principle. Also, with the aid of such geometry-matter coupling gravity, it is possible to avoid the big-bang singularity \cite{frlm1}.  Energy conditions in $\mathpzc{f}(\mathcal{R},\mathscr{L}_m)$ gravity studied in \cite{frlm2}. Numerous works are carried out to explain the cosmology in $\mathpzc{f}(\mathcal{R},\mathscr{L}_m)$ gravity \cite{frlm3,frlm4, frlm5}. A thermodynamic point of view is provided in \cite{frlmthermo}. In this work, we investigate traversable wormhole geometry in the context of $\mathpzc{f}(\mathcal{R},\mathscr{L}_m)$ gravity.
	
	\par The present manuscript is organized as follows: Section \ref{II} provides field equations in $\mathpzc{f}(\mathcal{R},\mathscr{L}_m)$ gravity. In section \ref{III}, we discuss the criteria for a traversable wormhole.  In section \ref{IV}, wormhole geometry in $\mathpzc{f}(\mathcal{R},\mathscr{L}_m)$ can be seen. Further, we explore different wormhole models in section \ref{V}. Finally, the last section \ref{VI} gives the discussion of the result and concluding remark.

	%%%%%%%%%%%%%%%%%%%%%%%%%%%%%%%%%%%%%%%%%%%%%%%%%%%%%%%%%%%%%%%%%%%%%%%%%%%
	\section{THE FIELD EQUATIONS IN $\mathpzc{f}(\mathcal{R},\mathscr{L}_m)$ GRAVITY}\label{II}			
		\par A novel approach of a modified gravity theory to deal with the governing field equations is the generalization of the action describing them. The action for $\mathpzc{f}(\mathcal{R},\mathscr{L}_m)$ gravity is given by,
		
		\begin{equation}\label{action}
			S=\int \mathpzc{f}(\mathcal{R},\mathscr{L}_m)	\sqrt{-g}\, d^4x,
		\end{equation}
		where, $\mathpzc{f}$ represents an arbitrary function of scalar curvature $\mathcal{R}$ and the matter lagrangian $\mathscr{L}_m$. For $\mathpzc{f}=\mathcal{R}/2+\mathscr{L}_m$, one can retain the governing equations of GR.
		\par The variation of \eqref{action} with respect to the metric tensor $g^{\mu\nu}$ provides the field equation for $\mathpzc{f}(\mathcal{R},\mathscr{L}_m)$ gravity that reads,
		\begin{equation}\label{fieldequation1}
			\begin{split}
				\mathpzc{f}_\mathcal{R}\mathcal{R}_{\mu\nu}+(g_{\mu\nu}\nabla_\mu\nabla^{\mu}-\nabla_\mu\nabla_\nu)\mathpzc{f}_\mathcal{R}-\dfrac{1}{2}\left[\mathpzc{f}- \mathpzc{f}_{\mathscr{L}_m}\mathscr{L}_m\right]g_{\mu\nu}\\=\dfrac{1}{2}\mathpzc{f}_{\mathscr{L}_m}\mathcal{T}_{\mu\nu}.
			\end{split}
		\end{equation}
		Here, $\mathpzc{f}_{\mathscr{L}_m}\equiv\frac{\partial \mathpzc{f}}{\partial \mathscr{L}_m}$, $\mathpzc{f}_\mathcal{R}\equiv\frac{\partial \mathpzc{f}}{\partial \mathcal{R}}$ and $\mathcal{T}_{\mu\nu}$ is the Energy-Momentum tensor (EMT) that takes the form,
		\begin{equation}
			\mathcal{T}_{\mu\nu}=-\dfrac{2}{\sqrt{-g}} \dfrac{\delta(\sqrt{-g}\mathscr{L}_m)}{\delta g^{\mu\nu}}=g_{\mu\nu}\mathscr{L}_m-2\dfrac{\partial \mathscr{L}_m}{\partial g^{\mu\nu}}.
		\end{equation}  
		\par Further, from the explicit form of the gravitational field equation \eqref{fieldequation1}, the covariant divergence of EMT becomes,
		\begin{equation}\label{divofT}
			\nabla^\mu \mathcal{T}_{\mu\nu}=2\left\lbrace \nabla^\mu \text{ln}\left[\mathpzc{f}_{\mathscr{L}_m} \right]\right\rbrace \dfrac{\partial \mathscr{L}_m }{\partial g^{\mu\nu}}. 
		\end{equation}
		\par Moreover, on contracting \eqref{fieldequation1} we get,
		\begin{equation}\label{traceoffieldequation}
			\begin{split}
				3\nabla_\mu\nabla^{\mu}\mathpzc{f}_\mathcal{R}+\mathpzc{f}_\mathcal{R}\mathcal{R}-2\left[\mathpzc{f} -\mathpzc{f}_{\mathscr{L}_m}\mathscr{L}_m\right]=\dfrac{1}{2}\mathpzc{f}_{\mathscr{L}_m}\mathcal{T}.
			\end{split}
		\end{equation}
		This provides the relation between the trace of EMT $\mathcal{T}=\mathcal{T}^\mu_\mu$, matter Lagrangian density $\mathscr{L}_m$ and the Ricci scalar $\mathcal{R}$.
		
		\par Now, using equations \eqref{fieldequation1} and \eqref{traceoffieldequation}, we obtain another form of gravitational field equation for $\mathpzc{f}(\mathcal{R},\mathscr{L}_m)$ gravity and is given by, 
		\begin{equation}\label{fieldquation2}
			\begin{split}
				\mathpzc{f}_\mathcal{R}\left( \mathcal{R}_{\mu\nu}-\dfrac{1}{3}\mathcal{R}g_{\mu\nu}\right) + \dfrac{g_{\mu\nu}}{6}\left[\mathpzc{f} -\mathpzc{f}_{\mathscr{L}_m}\mathscr{L}_m\right]\\=\dfrac{1}{2}\left(\mathcal{T}_{\mu\nu} -\dfrac{1}{3}\mathcal{T}g_{\mu\nu}\right)\mathpzc{f}_{\mathscr{L}_m}(\mathcal{R},\mathscr{L}_m)+\nabla_\mu\nabla_{\nu}\mathpzc{f}_\mathcal{R}.
			\end{split}
		\end{equation}	

%%%%%%%%%%%%%%%%%%%%%%%%%%%%%%%%%%%% Introduction to Wormhole %%%%%%%%%%%%%%%%%%%%%%%%%%%%%%%%%%%%%%%%
    \section{CRITERIA FOR A TRAVERSABLE WORMHOLE}\label{III}
		\par  A spherically symmetric non-rotating  Morris-Thorne wormhole metric \cite{morrisandthorne} in the Schwarzschild coordinates $(t,r,\theta,\phi)$ is given by, 
		\begin{equation}\label{whmetric}
			ds^2=-e^{2\varphi(r)}dt^2+\dfrac{dr^2}{1-\dfrac{b(r)}{r}  }  + r^2\left(d\theta^2+\text{sin}^2\theta \,d\phi^2\right). 
		\end{equation}  
		Here, $\varphi(r)$ is the gravitational redshift function, and $b(r)$ is the shape function. To examine the traversability of wormhole we consider yet another function, $\mathit{\mathit{l}}(r)$, the proper radial distance function and is expressed as, 
		\begin{equation*}
			\mathit{\mathit{l}}(r)=\pm \int_{r_0}^r \dfrac{dr}{\sqrt{\dfrac{r-b(r)}{r}}}.
		\end{equation*} 
	    \par These functions should satisfy certain criteria for a wormhole to be traversable.
	    \textit{Radial coordinate $r$:} The radial coordinate $r$ should always be positive and its minimum value $r_0(>0)$ is the throat radius. So we have, $r_0\le r<\infty$. 
	    \textit{Gravitational redshift function $\varphi(r)$:} To avoid the existence of the event horizon, the value of $\varphi(r)$ should always be finite everywhere. Further, the nature of the derivative of this redshift function is so significant in determining the geometrical aspects of the wormhole. 
	    \textit{Proper radial distance function $l(r)$:}  This function should be finite over radial coordinates $r$. In magnitude, it decreases from the upper universe to the throat and then increases from the throat to the lower universe. 
	    \textit{Shape function $b(r)$:} the shape function $b(r)$ should obey the following conditions;
		\begin{enumerate}[label=$\circ$,leftmargin=*]
			\setlength{\itemsep}{4pt}
			\setlength{\parskip}{4pt}
			\setlength{\parsep}{4pt}
			\item \textit{Throat condition}: The value of the function $b(r)$ at the throat is $r_0$ and hence $1-\frac{b(r)}{r}>0$ for $r>r_0.$ 
			\item \textit{Flaring-out condition:} The radial differential of the shape function, $b'(r)$ at the throat should satisfy, $b'(r_0)<1.$ 
			\item \textit{Asymptotic Flatness condition:} As $r\rightarrow \infty$, $\frac{b(r)}{r}\rightarrow 0$.
		\end{enumerate}

		%%%%%%%%%%%%%%%%%%%%%%%%%%%%%%%%%%%% Wormhole Solutions in f(R,\mathscr{L}_m) Gravity %%%%%%%%%%%%%%%%%%%%%%%%%%%%%%%%%%%%%%%%	
\section{WORMHOLE SOLUTIONS IN $\mathpzc{f}(\mathcal{R},\mathscr{L}_m)$ GRAVITY}\label{IV}
		
		\par In this work, for the traversability of the wormhole, the gravitational redshift function is supposed to be a constant, i.e., $\varphi'(r)=0$ or in other words, we are considering the wormhole having no tidal force. Also, the matter distribution is presumed to be anisotropic. So, EMT can be written as,
		\begin{equation}\label{energymomentumtensor}
			\mathcal{T}_{\mu\nu}=(\rho+p_t)\mathit{U}_\mu \mathit{U}_\nu-p_t\,g_{\mu\nu}+(p_r-p_t)\mathit{X}_{\mu}\mathit{X}_\nu,
		\end{equation}
		where $\rho$, $p_r$, and $p_t$ are respectively the energy density, the radial pressure, and the tangential pressure. Here, $\mathit{U}_\mu$ denotes a four-velocity vector with the unit norm, and $\mathit{X}_\mu$ denotes a space-like unit vector. Further, in this case, the tangential pressure will be orthogonal to $\mathit{X}_\mu$ and the radial pressure $p_r$ will be along $\mathit{U}_\mu$.
		
		\par Now, for the  Morris and Thorne wormhole metric \eqref{whmetric} with anisotropic matter distribution \eqref{energymomentumtensor}, the field equation \eqref{fieldquation2} can be depicted as,
		\begin{widetext}
		\begin{eqnarray}
		    \label{fe1}4\mathpzc{f}_\mathcal{R}\dfrac{b'}{r^2}-(\mathpzc{f}-\mathpzc{f}_{\mathscr{L}_m}\mathscr{L}_m)=(2\rho+p_r+2p_t)\mathpzc{f}_{\mathscr{L}_m}\\
		    \label{fe2}	6\mathpzc{f}_\mathcal{R}''\left(1-\dfrac{b}{r} \right)+3\mathpzc{f}_\mathcal{R}'\left(\dfrac{b-rb'}{r^2} \right)+2\mathpzc{f}_\mathcal{R}\left(\dfrac{3b-rb'}{r^3} \right) 
				-(\mathpzc{f}-\mathpzc{f}_{\mathscr{L}_m}\mathscr{L}_m)=(-\rho-2p_r+2p_t)\mathpzc{f}_{\mathscr{L}_m} \\
			\label{fe3}6\dfrac{\mathpzc{f}_\mathcal{R}''}{r}\left(1-\dfrac{b}{r} \right)-\mathpzc{f}_\mathcal{R}\left(\dfrac{3b-rb'}{r^3} \right)-(\mathpzc{f}-\mathpzc{f}_{\mathscr{L}_m}\mathscr{L}_m)=(-\rho+p_r-p_t)\mathpzc{f}_{\mathscr{L}_m} 
		\end{eqnarray}
		\end{widetext}
%%%%%%%%%%%%%%%%%%%%%%%%%%%%%%%%%%%% Energy Conditions %%%%%%%%%%%%%%%%%%%%%%%%%%%%%%%%%%%%%%%%
		
		\begin{center}
		    Energy Conditions
		\end{center}
		
		\par  Energy conditions determine the physical behavior of the motion of matter and energy that arise as a consequence of the Raychaudhuri equation. The studies on energy conditions in $\mathpzc{f}(\mathcal{R},\mathscr{L}_m)$ gravity can be seen in \cite{frlm2}. To analyze the geodesic behavior, we shall consider the criterion for different energy conditions. For the anisotropic matter distribution \eqref{energymomentumtensor} with $\rho$, $p_r$ and $p_t$ respectively being energy density, radial pressure, and tangential pressure, suppose that $n^\mu$ is a null vector and $u^\mu$ is a timelike vector, then, we have the following:
		\begin{enumerate}[label=$\circ$,leftmargin=*]
			\setlength{\itemsep}{4pt}
			\setlength{\parskip}{4pt}
			\setlength{\parsep}{4pt}
			\item \textit{Null Energy Conditions (NECs)}: Both $\rho+p_t$ and $\rho+p_r$ are non negative.
			\item \textit{Weak Energy Conditions (WECs)}: For non negative energy density, it implies $\rho+p_t$ and $\rho+p_r$ are both non negative.
			\item \textit{Strong Energy Conditions (SECs)}: For non negative $\rho+p_j$,  $\rho+\sum_j p_j$ is non negative $   \ \forall\ j$.
			\item \textit{Dominant Energy Conditions (DECs)}: For non negative energy density, it implies $\rho-|p_r|$ and $\rho-|p_t|$ are both non negative.
		\end{enumerate}

		\section{Wormhole Models}\label{V}
		\par In this section, we study non-minimal coupling of $\mathpzc{f}(\mathcal{R},\mathscr{L}_m)$ via two different cases. Primarily, we consider a specific form of the energy density function and secondly, we presume a particular shape function. In addition, we analyze energy conditions for both wormhole models.  
		
		\begin{center}
		    \textbf{Wormhole Model A}
		\end{center}
		\par For our present analysis, let us focus on the non-minimal coupling between $\mathcal{R}$ and $\mathscr{L}_m$ given by the model,
		\begin{equation}\label{mod}
		    \mathpzc{f}(\mathcal{R},\mathscr{L}_m)=\dfrac{\mathcal{R}}{2}+(1+\alpha \mathcal{R})\mathscr{L}_m
		\end{equation}
	    where $\alpha$ is a coupling constant and it determines the strength of coupling between scalar curvature and the matter Lagrangian. If $\alpha=0$, we can retain the field equations for GR. The main motivation for the choice of this model is the existence of curvature-matter coupling that arises from the term $\alpha\mathcal{R}\mathscr{L}_m$. In the seminal work of Harko, a general version of non-minimal coupling has been proposed (see ref \cite{model} for more details). It is inspired by the form, $\mathpzc{f}(\mathcal{R},\mathscr{L}_m) = f_1(\mathcal{R})+f_2(\mathcal{R})G(\mathscr{L}_m)$, where $f_1$ and $f_2$ are functions of RicciScalar and $G(\mathscr{L}_m)$ is a function of matter lagrangian. The current model is presenting the scenario for $f_1(\mathcal{R})=\frac{\mathcal{R}}{2}, f_2(\mathcal{R})=1+\alpha \mathcal{R}$ and $G(\mathscr{L}_m)=\mathscr{L}_m$ \cite{frlm,frlmmodel}.

     For the model in hand, the covariant derivative of EMT \eqref{divofT} becomes,
	    \begin{equation}
	        \nabla^\mu \mathcal{T}_{\mu\nu}=\dfrac{\alpha}{1+\alpha \mathcal{R}}\left( g_{\mu\nu}\mathscr{L}_m-\mathcal{T}_{\mu\nu}\right) \nabla^\mu \mathcal{R}
	    \end{equation}
	    \par In addition, in this case, we choose matter Lagrangian density as $\mathscr{L}_m=-\rho$. This choice for $\mathscr{L}_m$ has been extensively used in the literature (Readers may refer to \cite{whfrlm1,lmrho1,lmrho2,lmrho3,lmrho4}). Now, solving \eqref{fe1} and \eqref{fe2} for $p_r$ and $p_t$, we get, 
		\begin{widetext}
		\begin{align}
		    \label{pr}p_r&=\frac{2 \alpha r^{2}  (r-b)\rho''+(-\alpha r^{2}  \rho'-4 r \alpha \rho+r) b'-\rho r^{3}+\alpha b\rho' r+2\alpha b\rho -b}{r (2 \alpha b'+r^{2})}\\
		    \label{pt}p_t&=\frac{-2 \alpha r^{2}  (r-b) \rho''+r \alpha (r \rho'-4 \rho) b'-\rho r^{3}-\alpha b \rho'   r-2\alpha b\rho  +b}{2 r (2 \alpha b'+r^{2})}
		\end{align}
		\end{widetext}
		\textit{ Specific energy density:} Here, a presumption is made on the energy density by choosing a specific power law function \cite{powerlaw1,powerlaw2} given by,
	    \begin{equation}\label{density}
	        \rho=\rho_0 \left(\dfrac{r_0}{r}\right)^n
	    \end{equation}   
	    where, $n>0$ and $\rho_0>0$ are some constants. This power-law form of energy density helps us to solve the field equation \eqref{fe3}. The shape function so obtained is given by,

	 \begin{widetext}
	 \begin{equation}\label{Asf}
	      \begin{split}
	         b(r)=\left[1+\alpha\rho_0(n-2) \left(\frac{r_0}{r}\right)^{n}\right]^{-\frac{6 n+7}{n-2}}\left[\rho_0\left(\frac{r_0}{r}\right)^{n} r \left\{6\alpha\frac{ n(n+1)}{n-1}\;{}_2F_1\left( 1-\frac{1}{n}, -\frac{5 n+9}{n-2}; 2-\frac{1}{n};-\alpha\rho_0(n-2)\left(\frac{r_0}{r}\right)^n \right)\right.\right.\\
	         \left.\left.-\frac{r^2}{n-3}\;{}_2F_1\left(-\frac{5 n+9}{n-2},\frac{n-3}{n};\frac{2n-3}{n} ;-\alpha\rho_0(n-2)\left(\frac{r_0}{r}\right)^n \right)\right\}+k\right],
	         \end{split}
	 \end{equation}

	  \end{widetext}
where, ${}_2F_1(a,b;c;z)$ is a hypergeometric function and $k$ is the constant of integration. Now, it is our task to validate the obtained shape function for traversability. So, as discussed earlier, $b(r)$ should satisfy the throat condition, flaring-out condition and asymptotically flatness condition. In order to satisfy the throat condition $b(r_0)=r_0$, $k$ takes the form,
    \begin{widetext}
    \begin{equation}
           \begin{split}
               k=r_0(1+\alpha \rho_0 (n-2))^{-\frac{6 n+7}{n-2}}-6 \alpha n r_0\rho_0 \frac{n+1}{n-1}  \;{}_2F_1\left( 1-\frac{1}{n}, -\frac{5 n+9}{n-2}; 2-\frac{1}{n} ;-\alpha\rho_0(n-2) \right)\\+\frac{r_0^3\rho_0}{n-3}\;{}_2F_1\left(\frac{n-3}{n},-\frac{5 n+9}{n-2};\frac{2n-3}{n} ;-\alpha\rho_0(n-2) \right).
           \end{split}
           \end{equation}
    \end{widetext}

    \begin{figure*}[!]
        \centering
        \subfloat[$b(r)>0, b(r)<r$ \label{fig:Asf1}]{\includegraphics[width=0.40\linewidth]{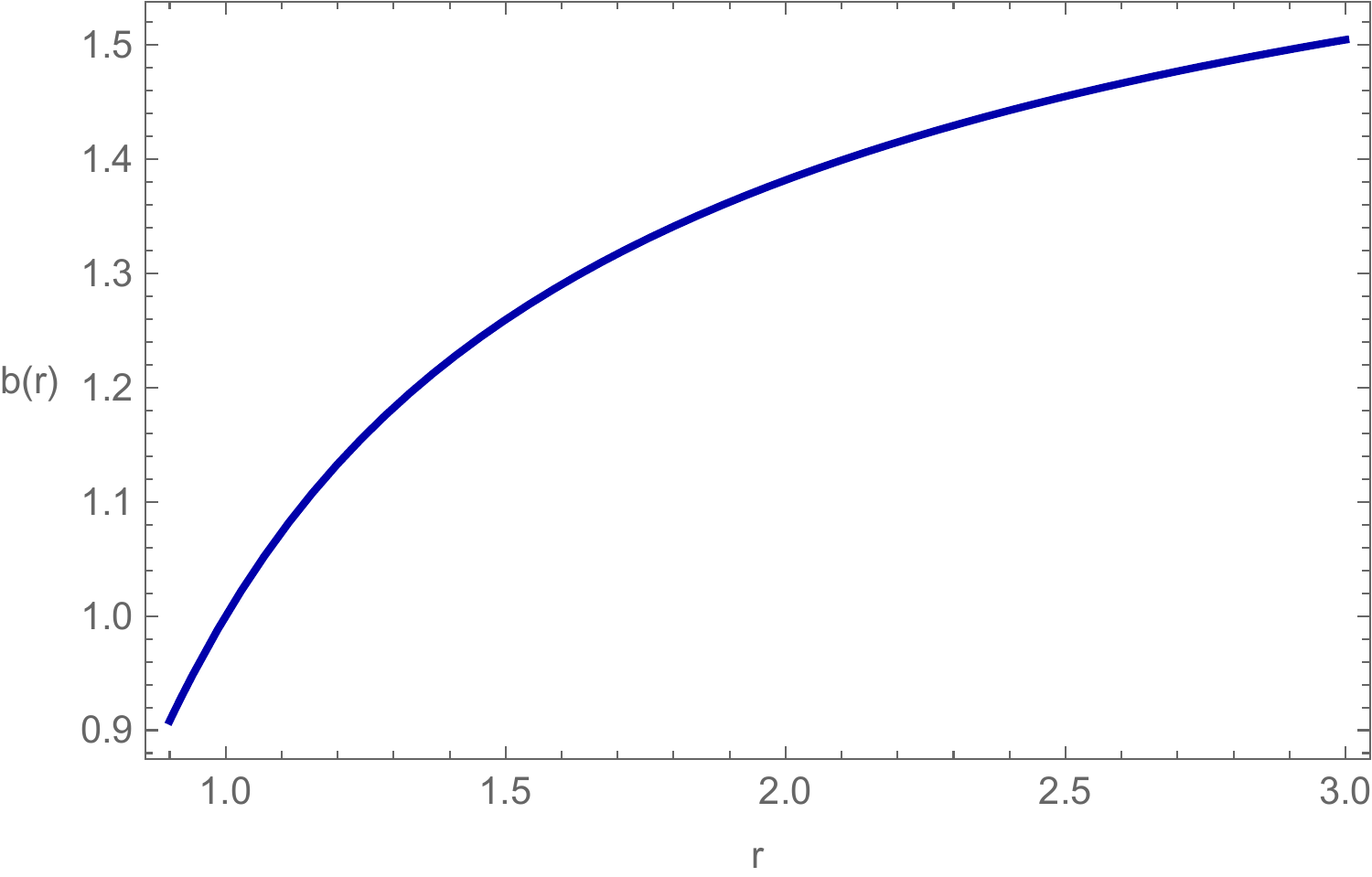}}
        \subfloat[$b'(r)<1$\label{fig:Asf2}]{\includegraphics[width=0.40\linewidth]{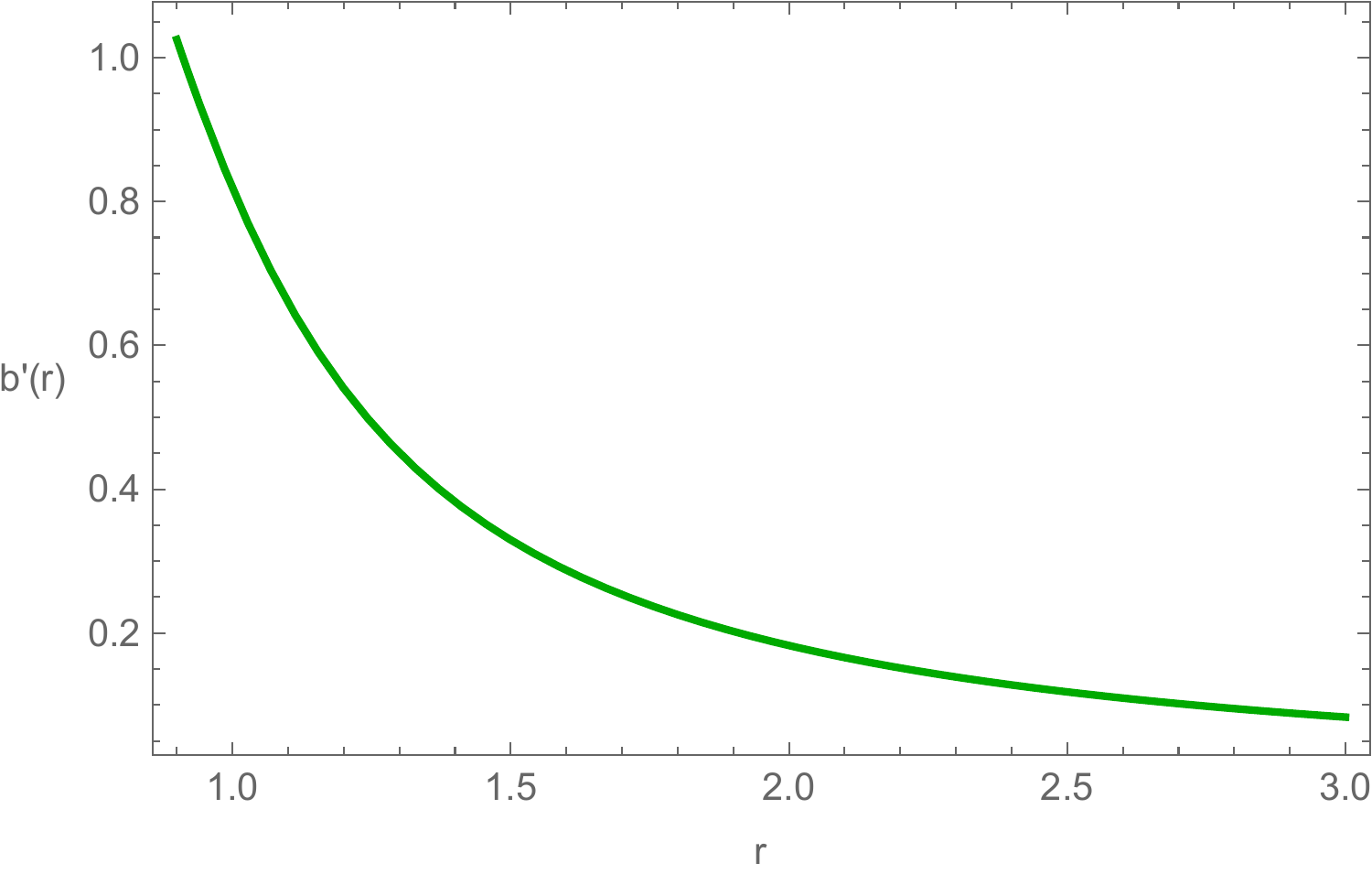}}\\
        \subfloat[$b(r)/r\to 0$ as $r\to \infty$\label{fig:Asf3}]{\includegraphics[width=0.40\linewidth]{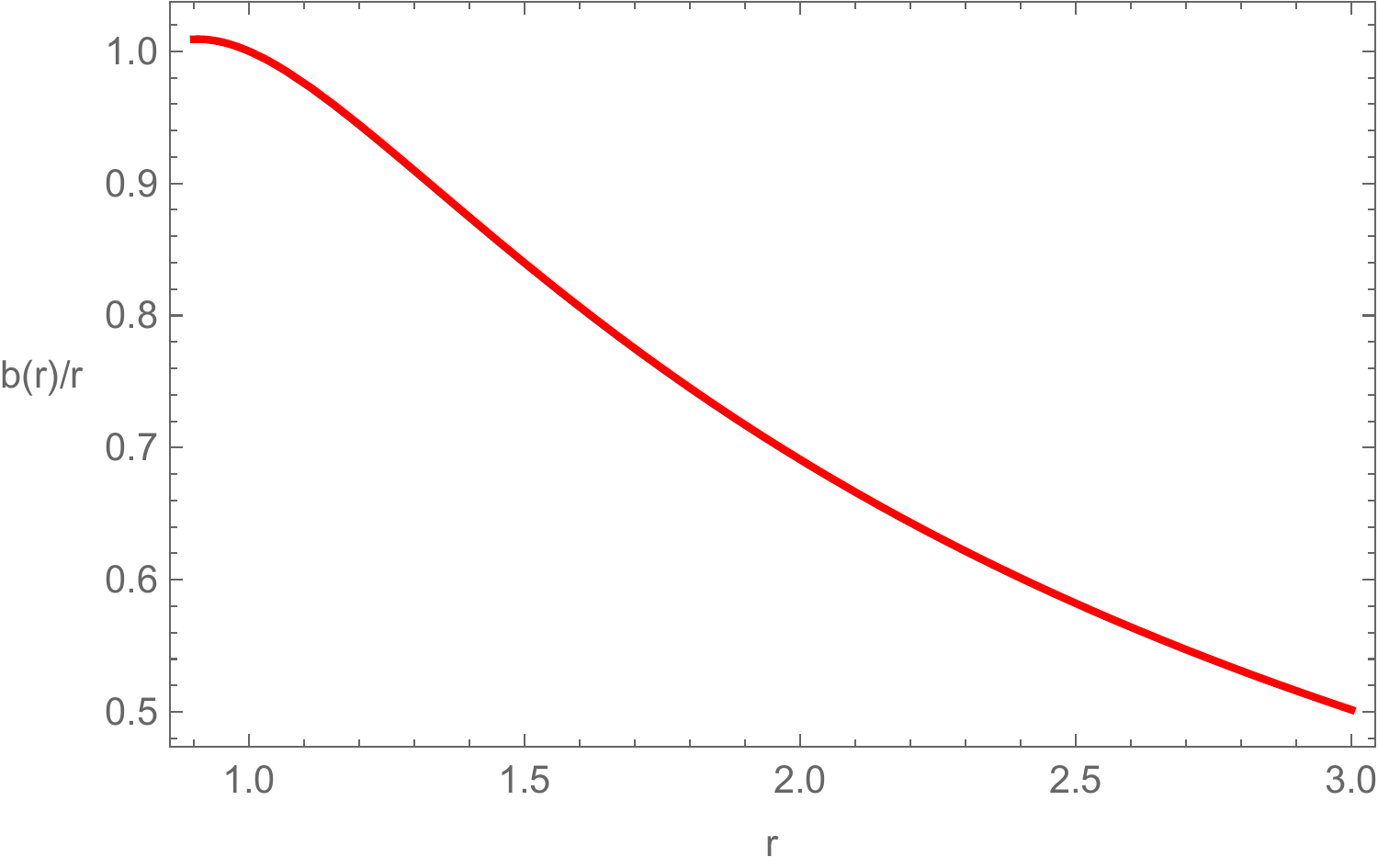}}
        \subfloat[$b(r)-r=0$ at $r=r_0$ \label{fig:Asf4}]{\includegraphics[width=0.40\linewidth]{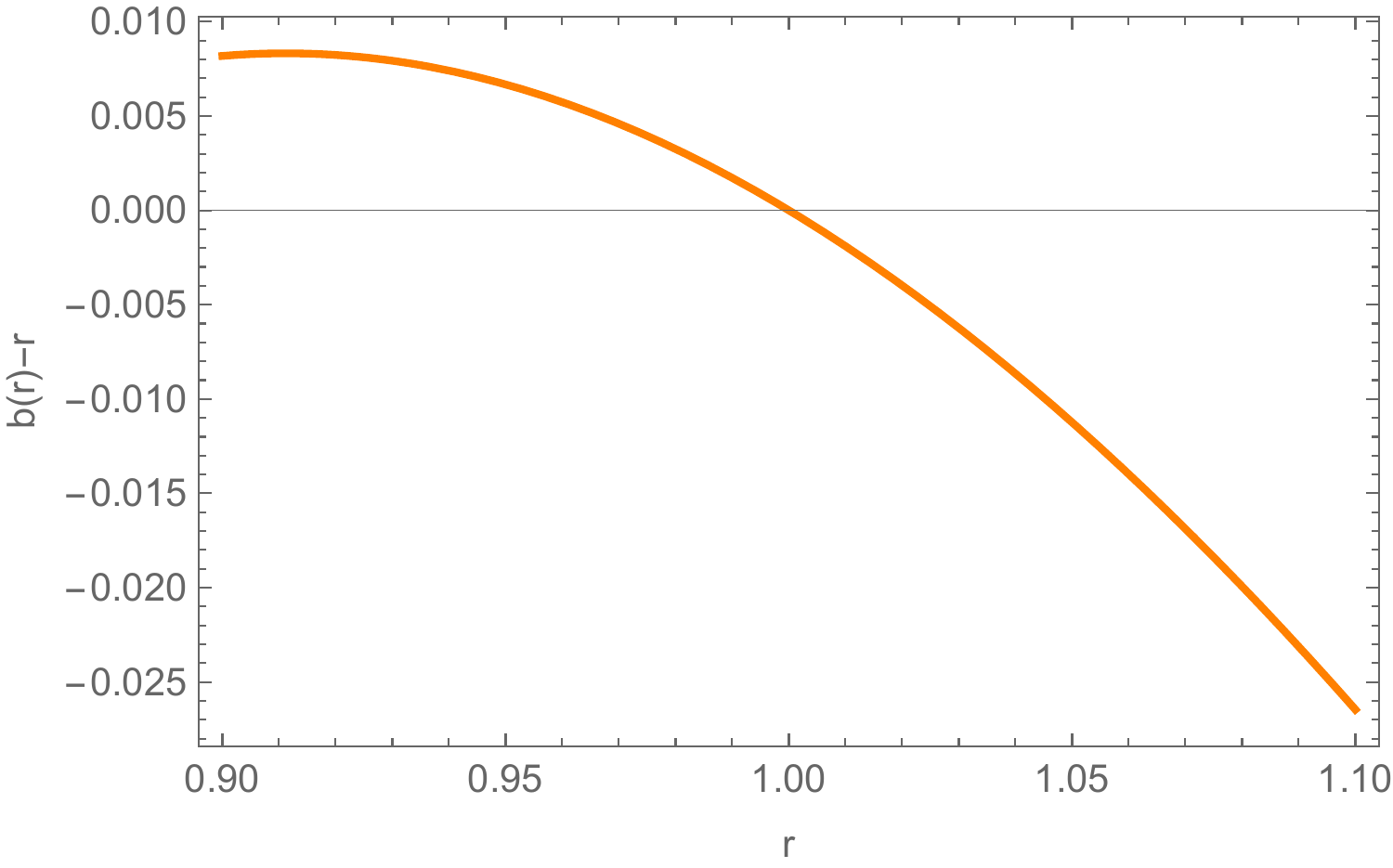}}\\
        \subfloat[$1-b(r)/r\to 1$ as $r\to\infty$\label{fig:Asf5}]{\includegraphics[width=0.40\linewidth]{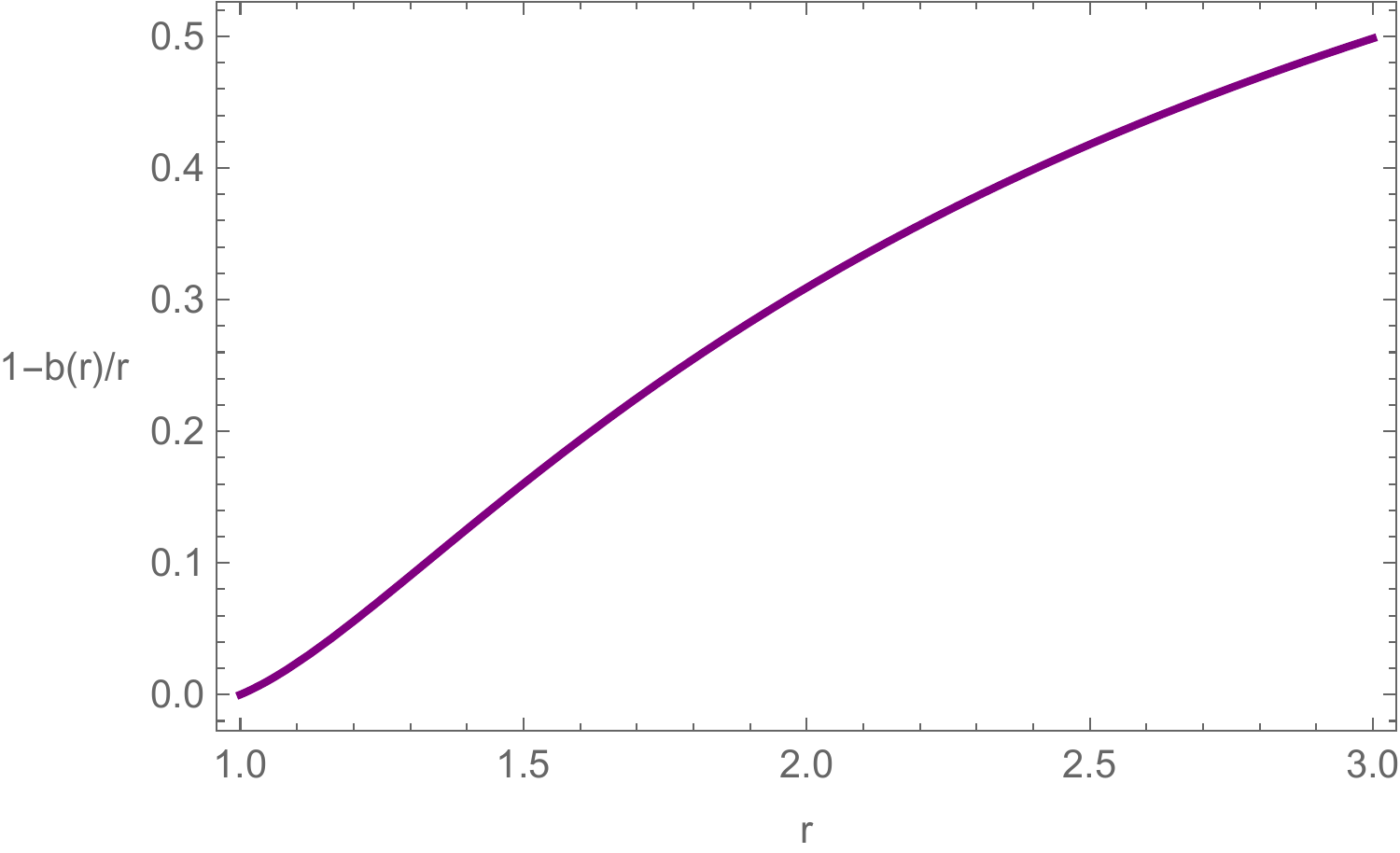}}
        \caption{Model A: A plot showing the characteristics of shape function $b$ with $r_0=1, \rho_0=0.8, \alpha=0.01$ and $n=4$. Here, for $r>r_0$ we have $b<r$, $b'<1$, $b/r\rightarrow 0$ as $r\rightarrow \infty$, $b-r=0$ at throat and $1-b(r)/r\to 1$ as $r\to\infty$.}
        \label{fig:Asf}
    \end{figure*}
    \par Further, the flaring-out condition at the throat $r=r_0$ for the shape function \eqref{Asf} is satisfied if the following constraining relation holds:
    \begin{equation*}\label{ineq}
        \rho_0(r_0^2+2\alpha)<1.
    \end{equation*} 
    Moreover, in the case of GR ($\alpha=0$), this relation gives $\rho_0<1/r_0^2$. Now, from the valid choice for the constants $r_0$ and $\rho_0$, we can constrain the model parameter $\alpha$. Say, for $\rho_0=0.8$ and $r_0=1$, we have, $\alpha<0.125$. In other words, for $\alpha<0.125$, $b(r)$ satisfies the flaring-out condition.

    Further, for the metric \eqref{whmetric}, the term $1-\frac{b(r)}{r}$ expresses the interpretation of horizon structure. For the shape function in hand, we have
    \begin{widetext}
        \begin{equation}
        \begin{split}
        1-\frac{b(r)}{r}=1-\frac{\left(1+\alpha\rho_0(n-2) \left(\frac{r_0}{r}\right)^{n}\right)^{-\frac{6 n+7}{n-2}}}{r}\left[ \left\{6\alpha\frac{ n(n+1)}{n-1}\;{}_2F_1\left( 1-\frac{1}{n}, -\frac{5 n+9}{n-2}; 2-\frac{1}{n};-\alpha\rho_0(n-2)\left(\frac{r_0}{r}\right)^n \right)\right.\right.\\
	         \left.\left.-\frac{r^2}{n-3}\;{}_2F_1\left(-\frac{5 n+9}{n-2},\frac{n-3}{n};\frac{2n-3}{n} ;-\alpha\rho_0(n-2)\left(\frac{r_0}{r}\right)^n \right)\right\}\rho_0\left(\frac{r_0}{r}\right)^{n} r+r_0(1+\alpha \rho_0 (n-2))^{-\frac{6 n+7}{n-2}}\right.\\\left.-6 \alpha n r_0\rho_0 \frac{n+1}{n-1}  \;{}_2F_1\left( 1-\frac{1}{n}, -\frac{5 n+9}{n-2}; 2-\frac{1}{n} ;-\alpha\rho_0(n-2) \right)+\frac{r_0^3\rho_0}{n-3}\;{}_2F_1\left(\frac{n-3}{n},-\frac{5 n+9}{n-2};\frac{2n-3}{n} ;-\alpha\rho_0(n-2) \right)\right].
        \end{split}
        \end{equation}
    \end{widetext}
   The fore-mentioned quantity is non-zero for the values of $r$ above the throat radius.
    
    \par In addition, it satisfies the asymptotic flatness condition for $n\ge4$.  \figureautorefname $\;$ \ref{fig:Asf} shows the characteristics of shape function $b(r)$ for $r_0=1, \rho_0=0.8, \alpha=0.01$ and $n=4$. Here, it can be observed that the shape function $b(r)$ is non-negative and increasing in the entire domain of radial coordinate $r$ [\figureautorefname$\;$\ref{fig:Asf1}]. Also, $b'(r)<1$ with $b'(1)\approx 0.8189$ [\figureautorefname$\;$\ref{fig:Asf2}] and $\lim_{r\to\infty} \frac{b(r)}{r}=0$ [\figureautorefname$\;$\ref{fig:Asf3}]. Further, $b(1)=1$ for $r=r_0=1$ [\figureautorefname$\;$\ref{fig:Asf4}] and $1-b(r)/r\to 1$ as $r\to\infty$ [\figureautorefname$\;$\ref{fig:Asf5}]. Thus, we can say that the derived shape function $b(r)$ satisfies all the necessary criteria. These results are summarized in \tableautorefname$\;$\ref{tab:table0}.
    
    Now, using equations \eqref{pr}-\eqref{density}, we can study the behavior of energy conditions [\figureautorefname $\;$\ref{fig:Aec}]. It can be seen that throughout the domain energy density remains positive [\ref{fig:Arho}]. Energy conditions aid us to analyse the nature of particle-energy motion. Here, for the present scenario, we can observe the violation of NEC and DEC for radial pressure. But these energy conditions are satisfied for $p_t$. Also, there is a violation of the SEC.   

    \begin{widetext}
	\begin{figure*}[t!]
	    \centering
	    \subfloat[Energy density $\rho$\label{fig:Arho}]{\includegraphics[width=0.31\linewidth]{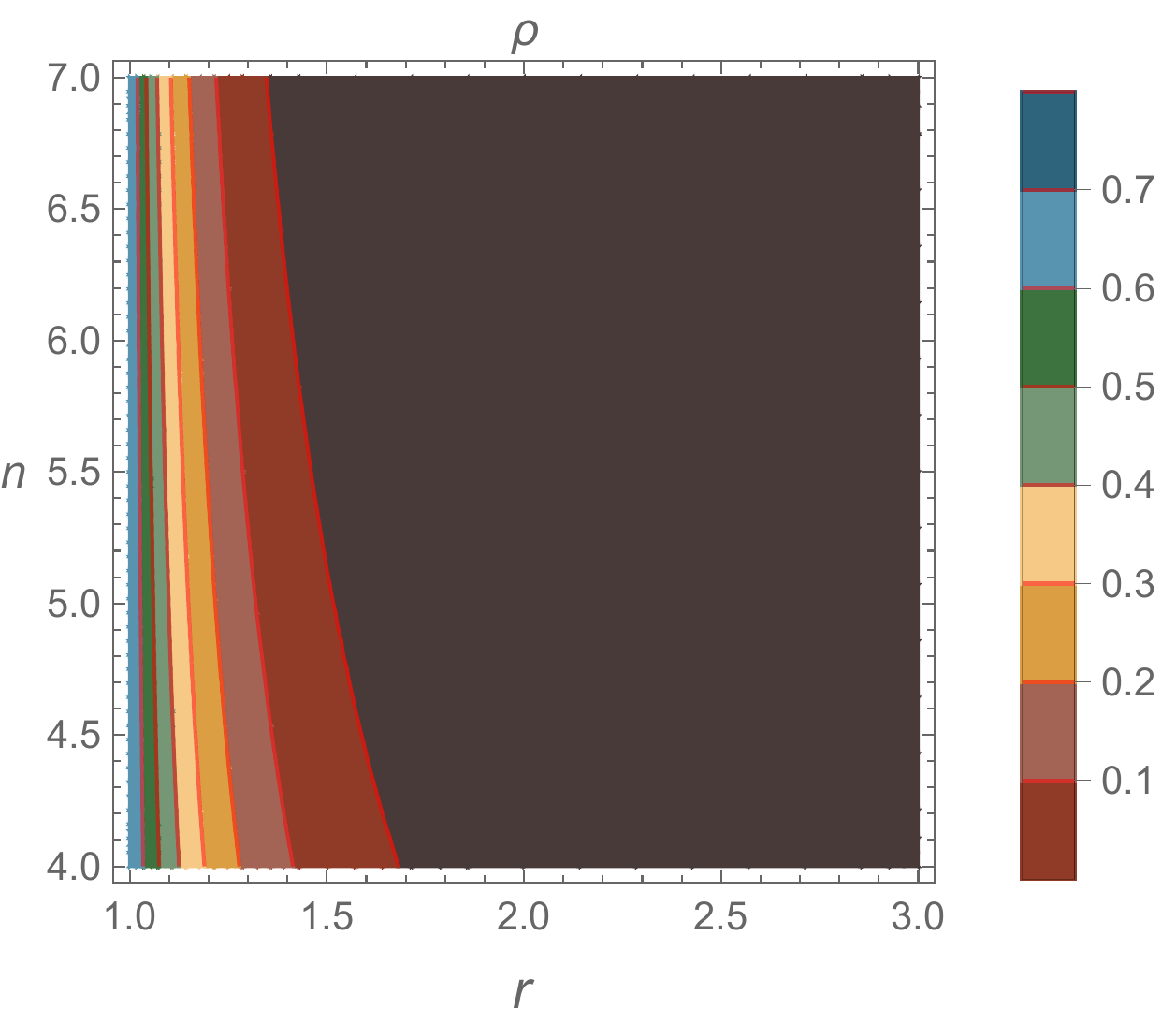}}
	    \subfloat[NEC $\rho+p_r$\label{fig:Ae1}]{\includegraphics[width=0.34\linewidth]{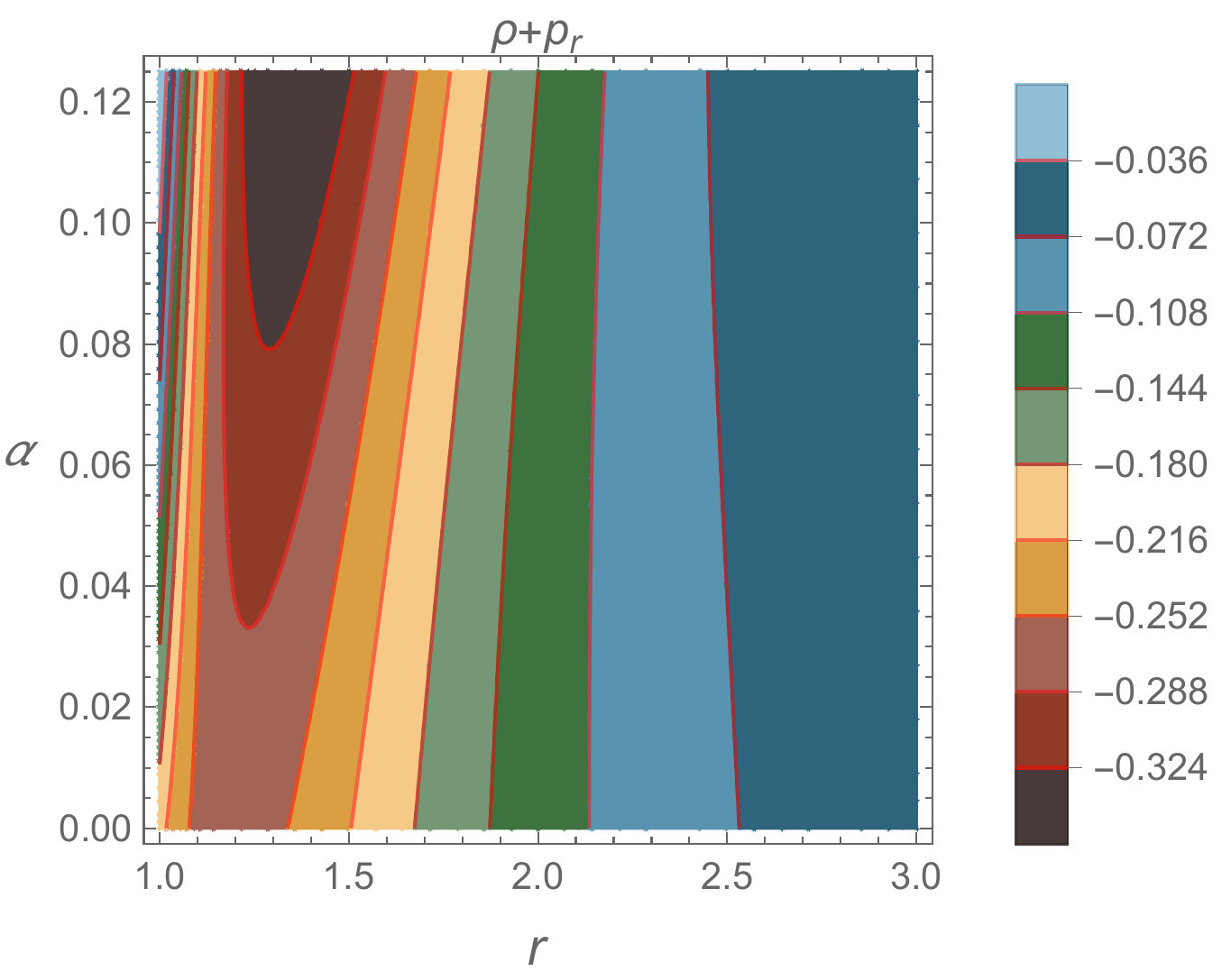}}
	    \subfloat[NEC $\rho+p_t$\label{fig:Ae2}]{\includegraphics[width=0.33\linewidth]{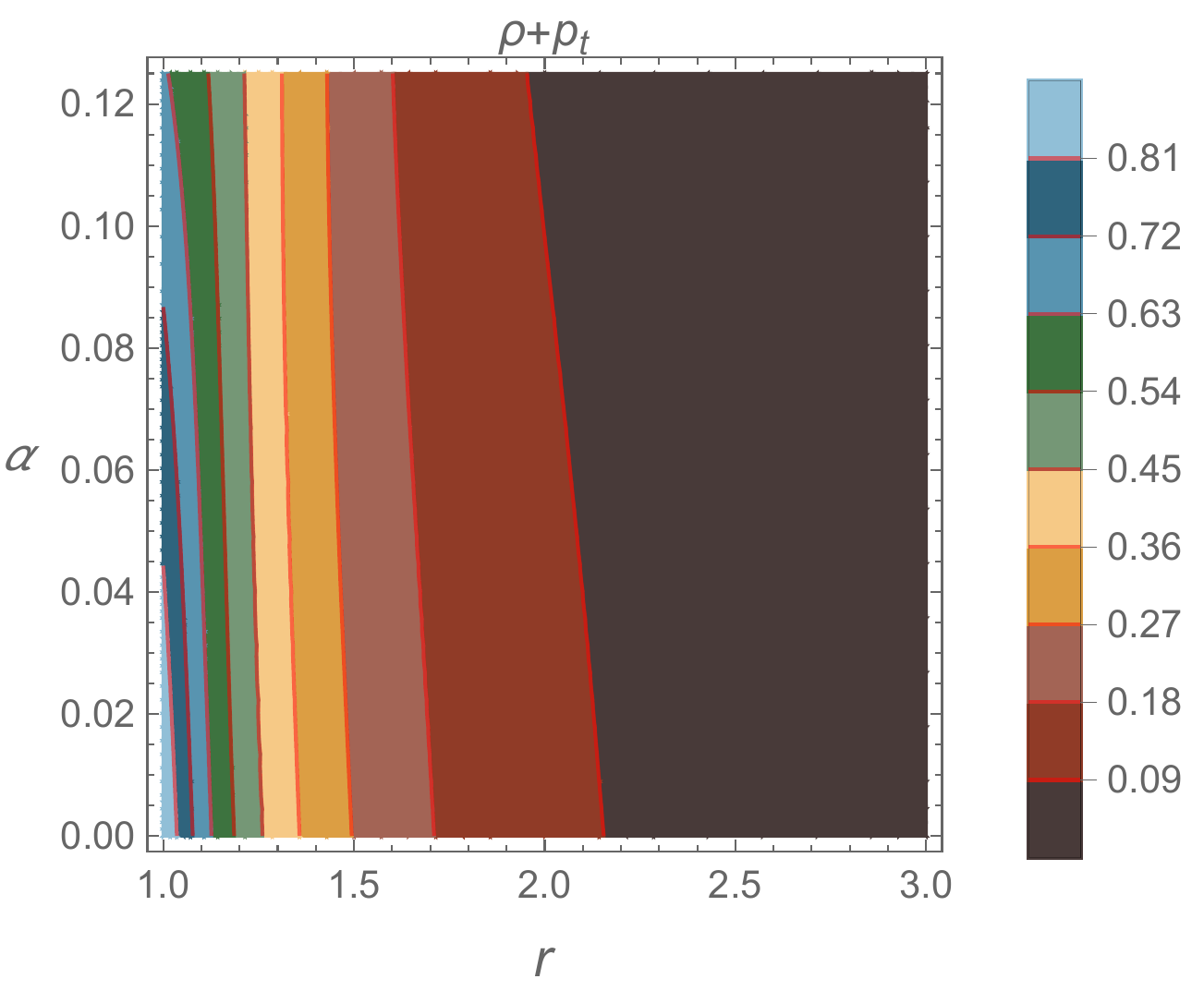}}\\
	    \subfloat[DEC $\rho-|p_r|$\label{fig:Ae3}]{\includegraphics[width=0.33\linewidth]{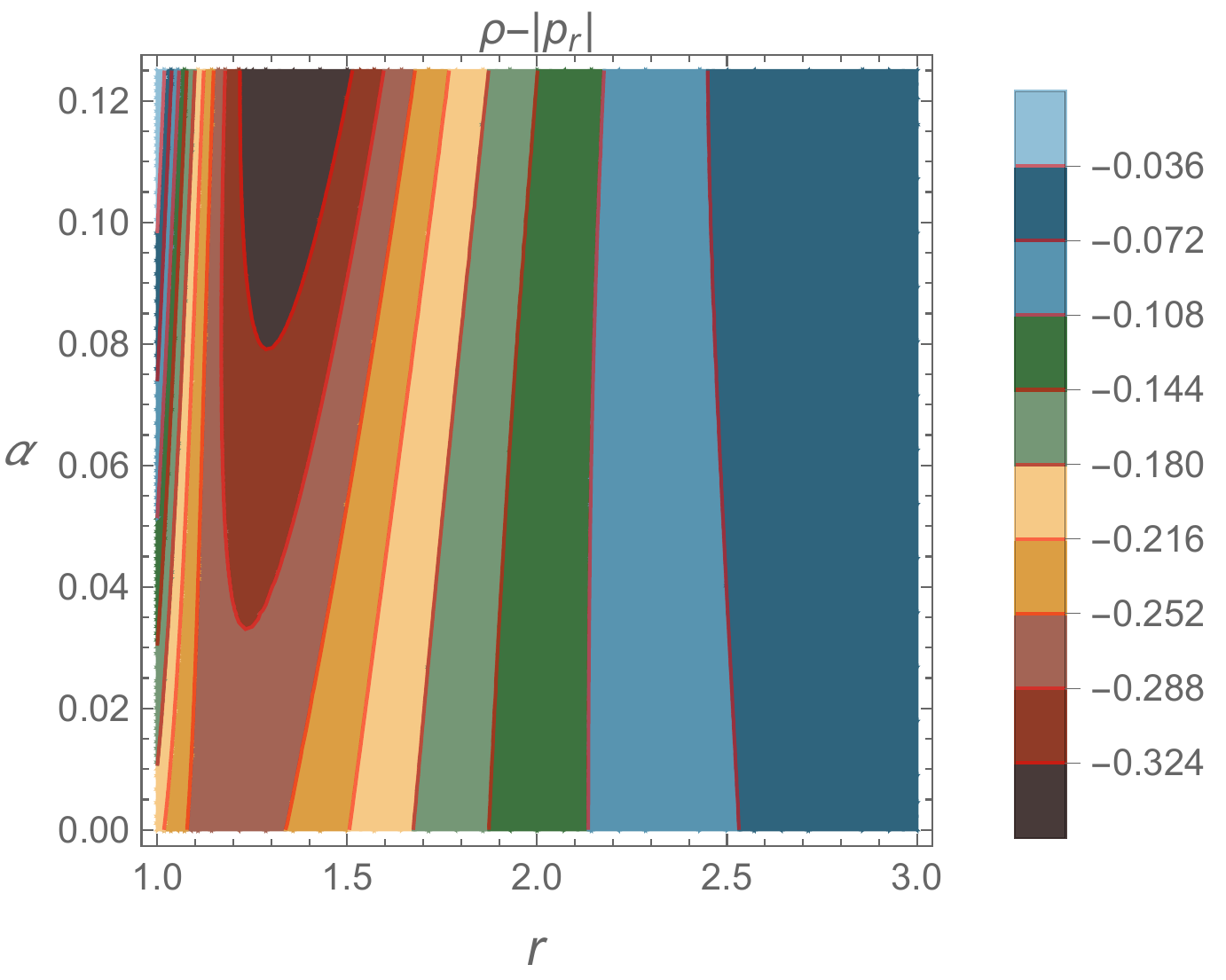}}
	    \subfloat[DEC $\rho-|p_t|$\label{fig:Ae4}]{\includegraphics[width=0.33\linewidth]{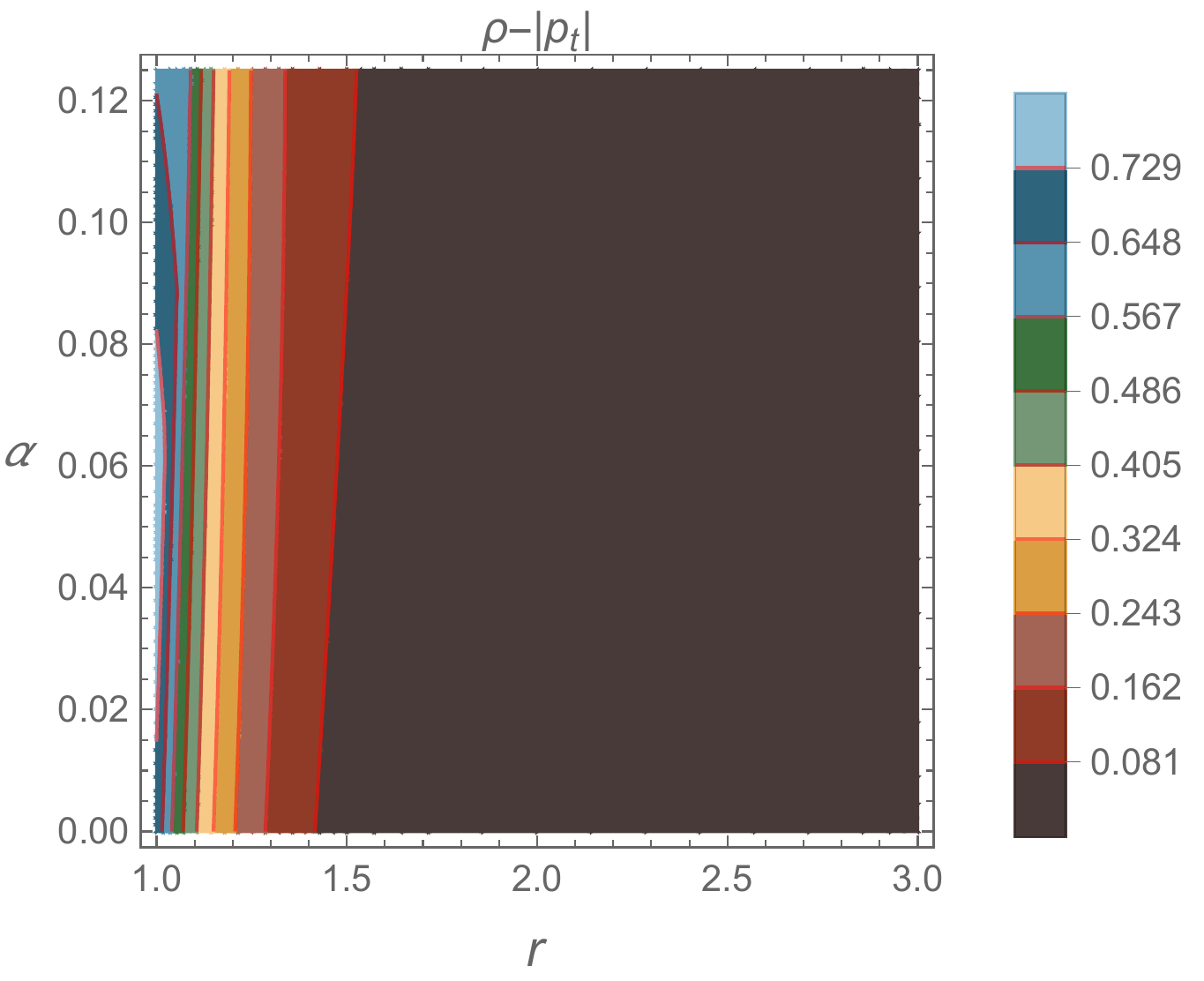}}
	    \subfloat[SEC $\rho+p_r+2p_t$\label{fig:Ae5}]{\includegraphics[width=0.33\linewidth]{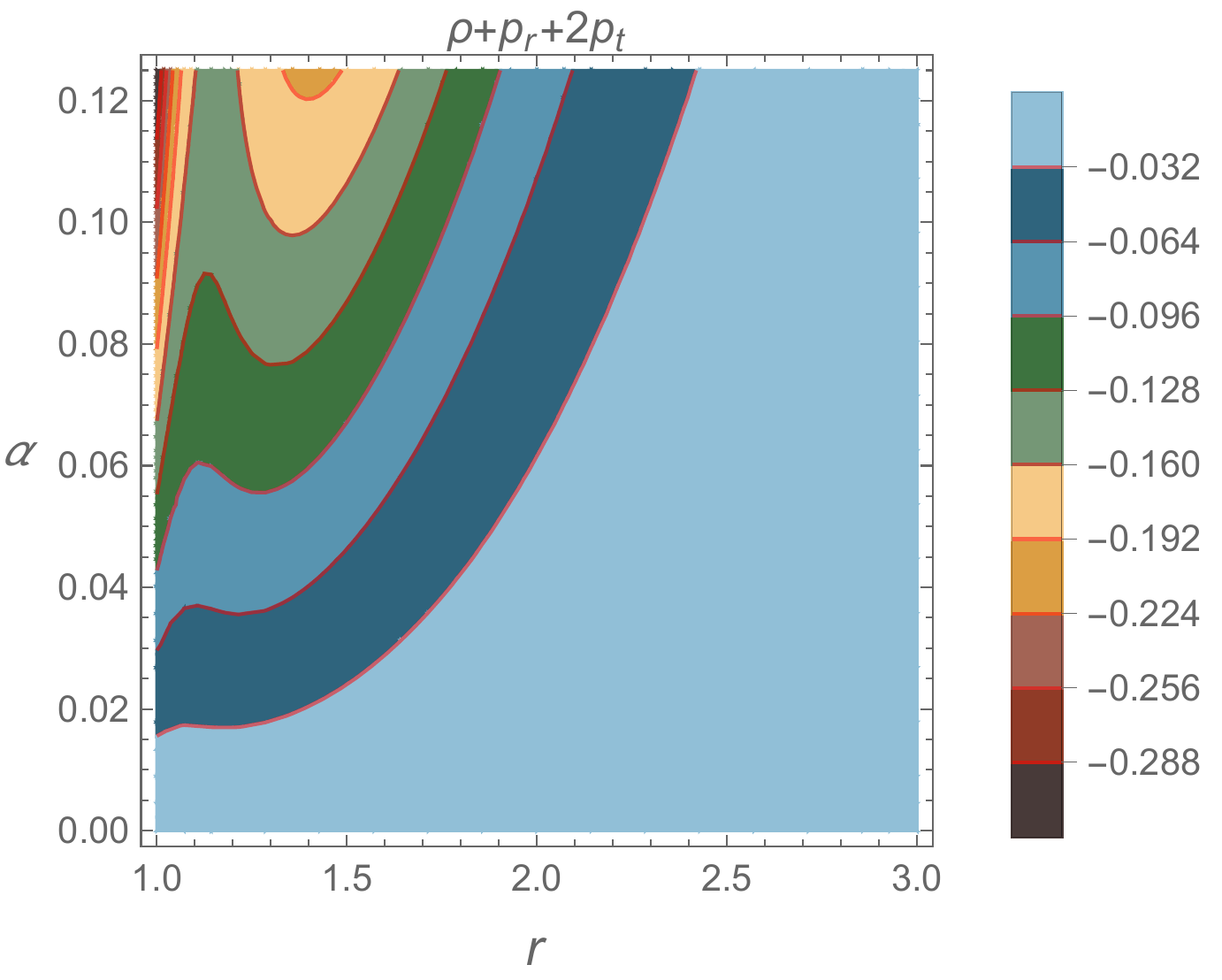}}
	    \caption{Model A: A plot depicting behaviour of (a) energy density with respect to $n$ for $\alpha=0.1, \rho_0=0.8$ and $r_0=1$ (b-f) energy conditions with respect to model parameter $\alpha$ for $n=4,\rho_0=0.8$ and $r_0=1$}
	    \label{fig:Aec}
	\end{figure*}
	
	 \begin{table*}[h!]
		\caption{Model A: Summarizing the nature of shape function.}
		    \label{tab:table0}
		%\begin{ruledtabular}
		    \centering
		    \begin{tabular}{|c|c|c|}
		        \hline
		        \textit{Function }           & \textit{Result}  & \textit{Interpretation} \\
		        \hline
		        $b(r)$  & \makecell{ $0<b(r)<r\;\forall\;r>r_0$ and  $\; b(r_0)=r_0$}  & \makecell{Viable form of shape function\\ and throat condition is satisfied}\\
		        \hline
		        $b'(r)$ & \makecell{$<1$, for $\alpha<0.125$,\\$\rho_0=0.8$ and $r_0=1$} & Flaring-out condition is satisfied \\
		       \hline
		       $\dfrac{b(r)}{r}$& \makecell{approaches to 0 for large value\\ of $r$ when $n\ge 4$ }& Asymptotic flatness condition is satisfied\\
		       \hline
                    $1-\frac{b(r)}{r}$ &\makecell{approaches to 1 for large value\\ of $r$ when $n\ge 4$}  & Horizon structure analysis\\
                    \hline
			\end{tabular}
		   %\end{ruledtabular}
		\end{table*}
		
	\begin{figure*}[!]
	    \centering
	    \subfloat[$b(r)>0, b(r)<r$ \label{fig:Bsf1}]{\includegraphics[width=0.40\linewidth]{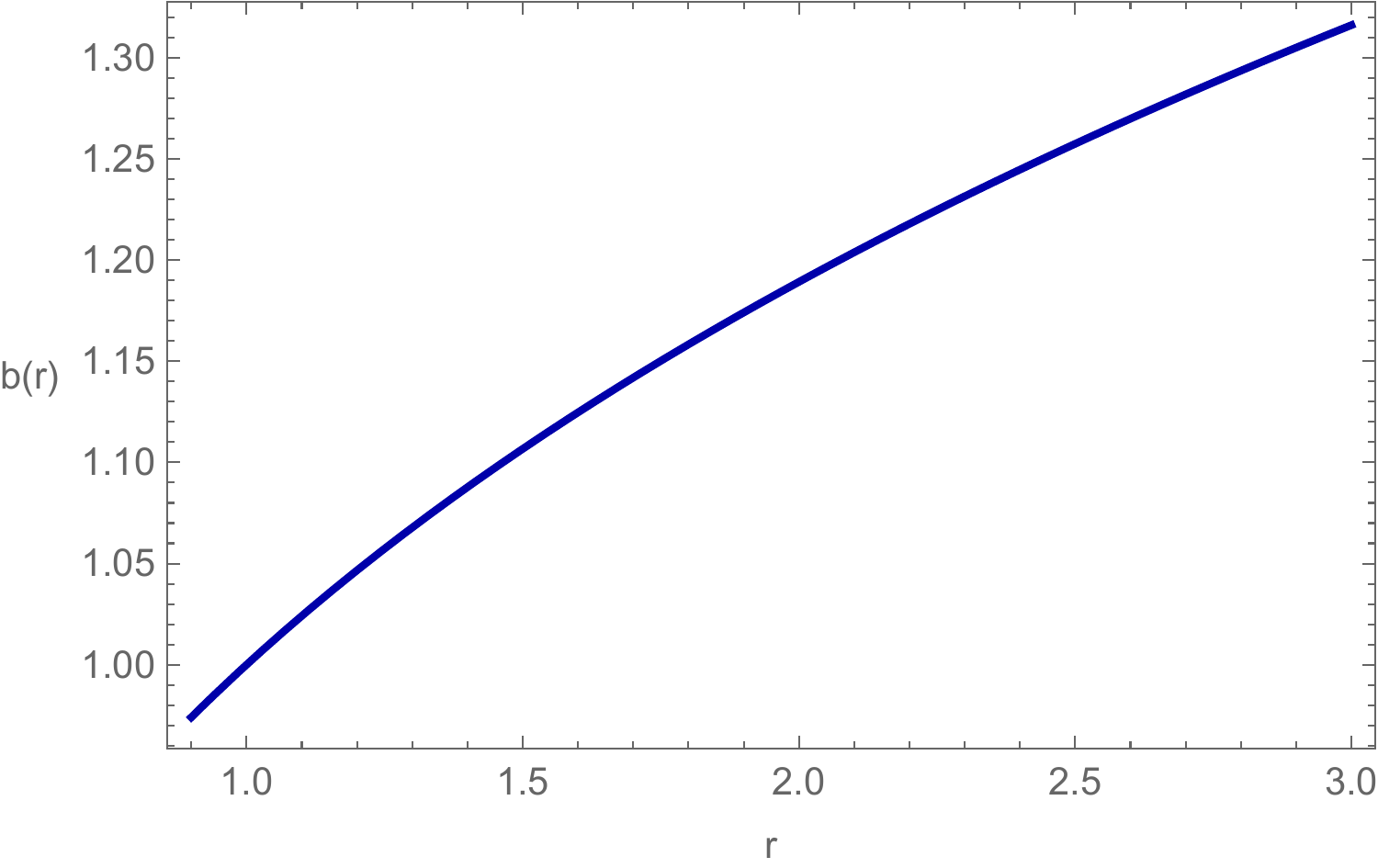}}
        \subfloat[$b'(r)<1$\label{fig:Bsf2}]{\includegraphics[width=0.40\linewidth]{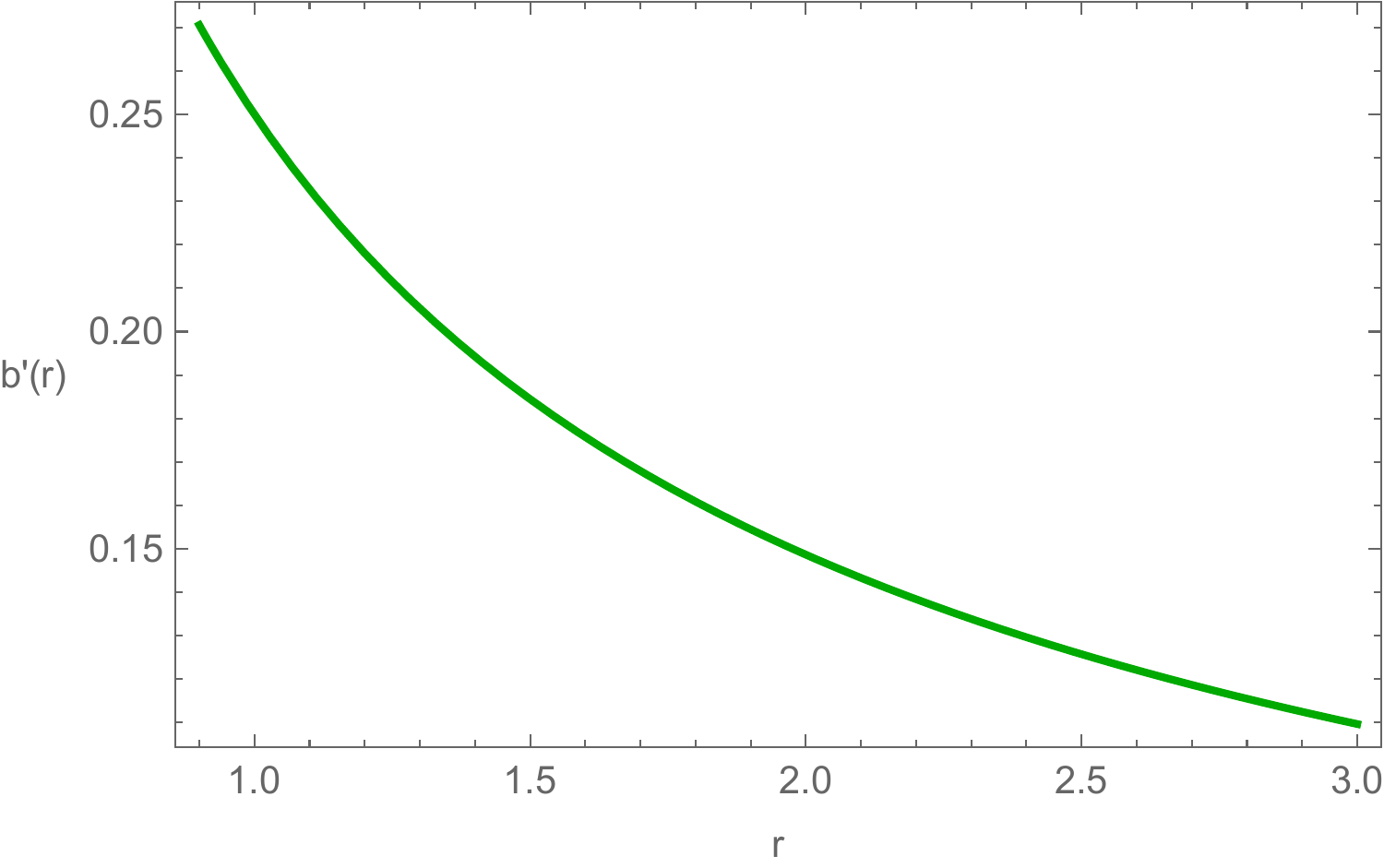}}\\
        \subfloat[$b(r)/r\to 0$ as $r\to \infty$\label{fig:Bsf3}]{\includegraphics[width=0.40\linewidth]{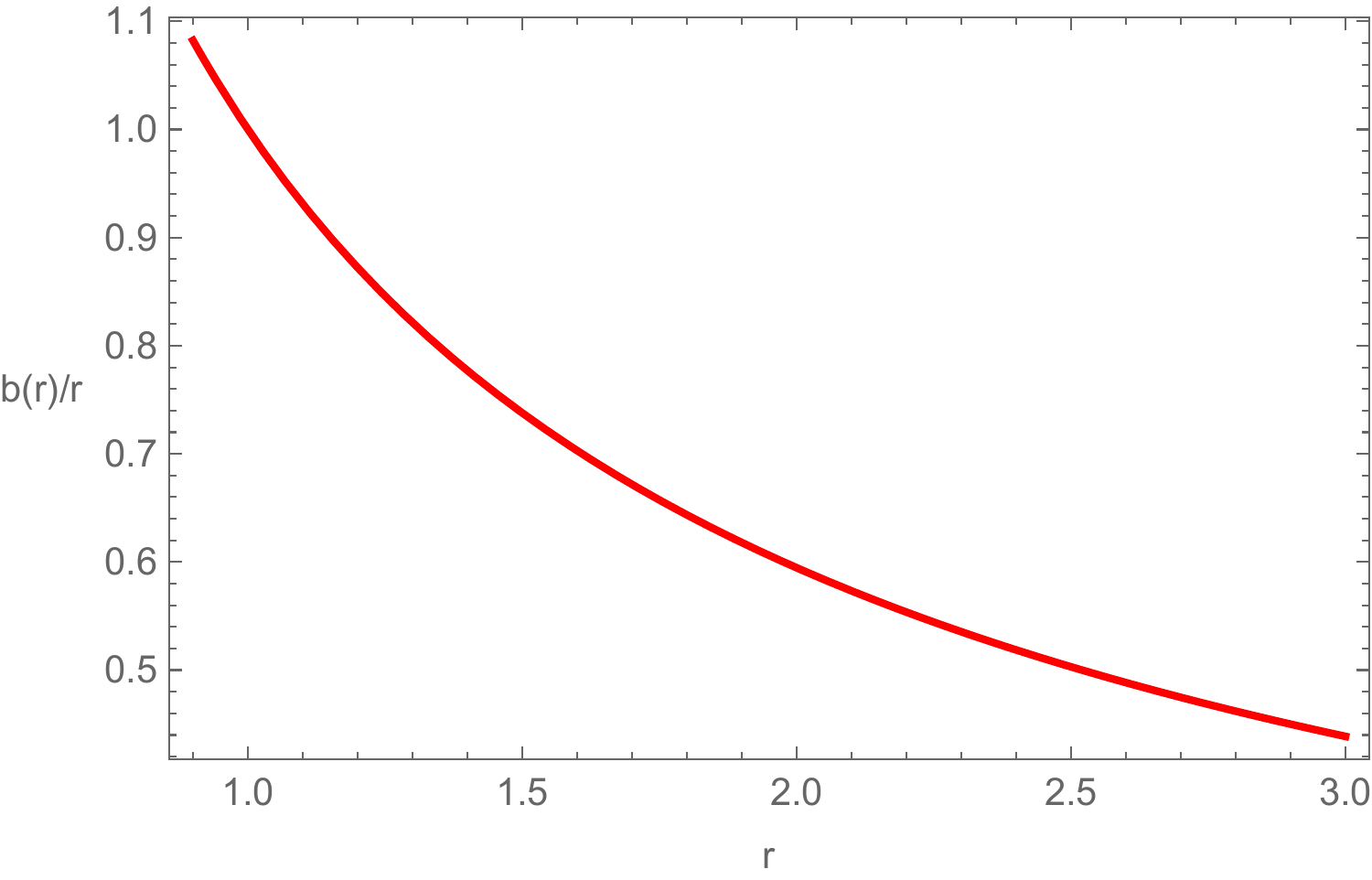}}
        \subfloat[$b(r)-r=0$ at $r=r_0$ \label{fig:Bsf4}]{\includegraphics[width=0.40\linewidth]{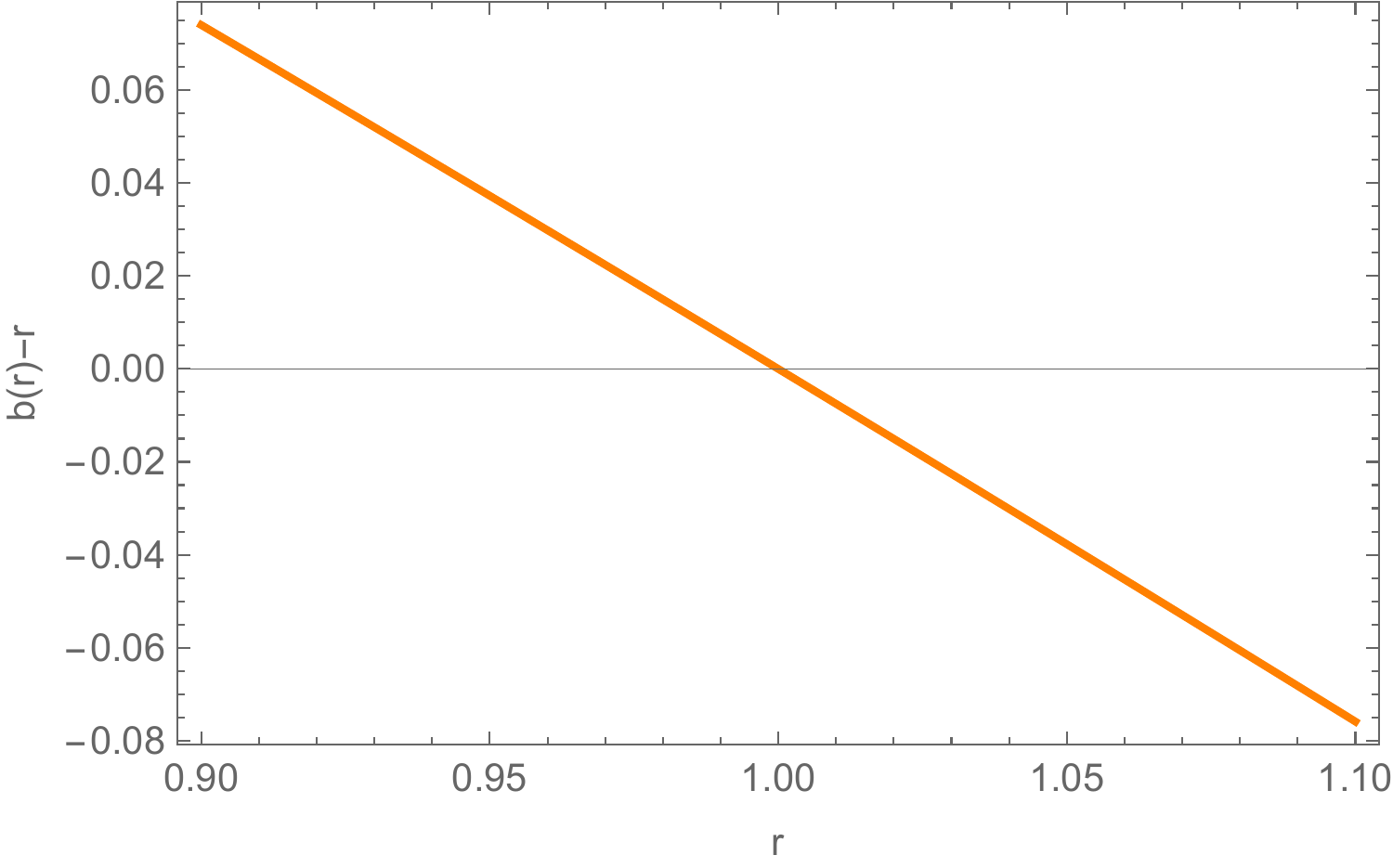}}\\
        \subfloat[$1-b(r)/r\to 1$ as $r\to\infty$ \label{fig:Bsf5}]{\includegraphics[width=0.40\linewidth]{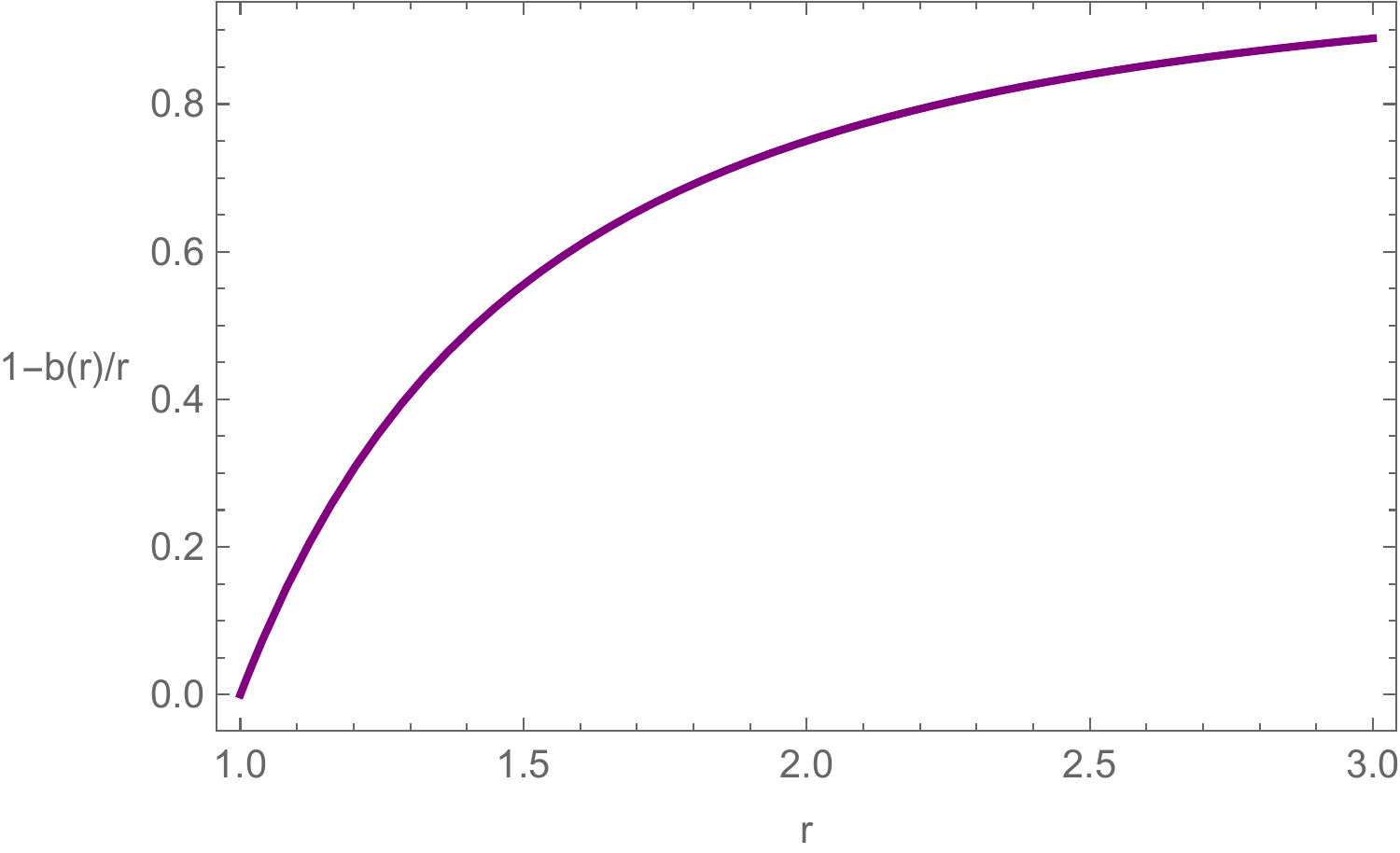}}
	    \caption{Model B: Profile of shape function for $m=0.25$ satisfying all the required conditions.}
	    \label{fig:Bsf}
	\end{figure*}
	
		\begin{figure*}[!]
	    \centering
	   \subfloat[Energy density $\rho$\label{fig:Brho}]{\includegraphics[width=0.4\linewidth]{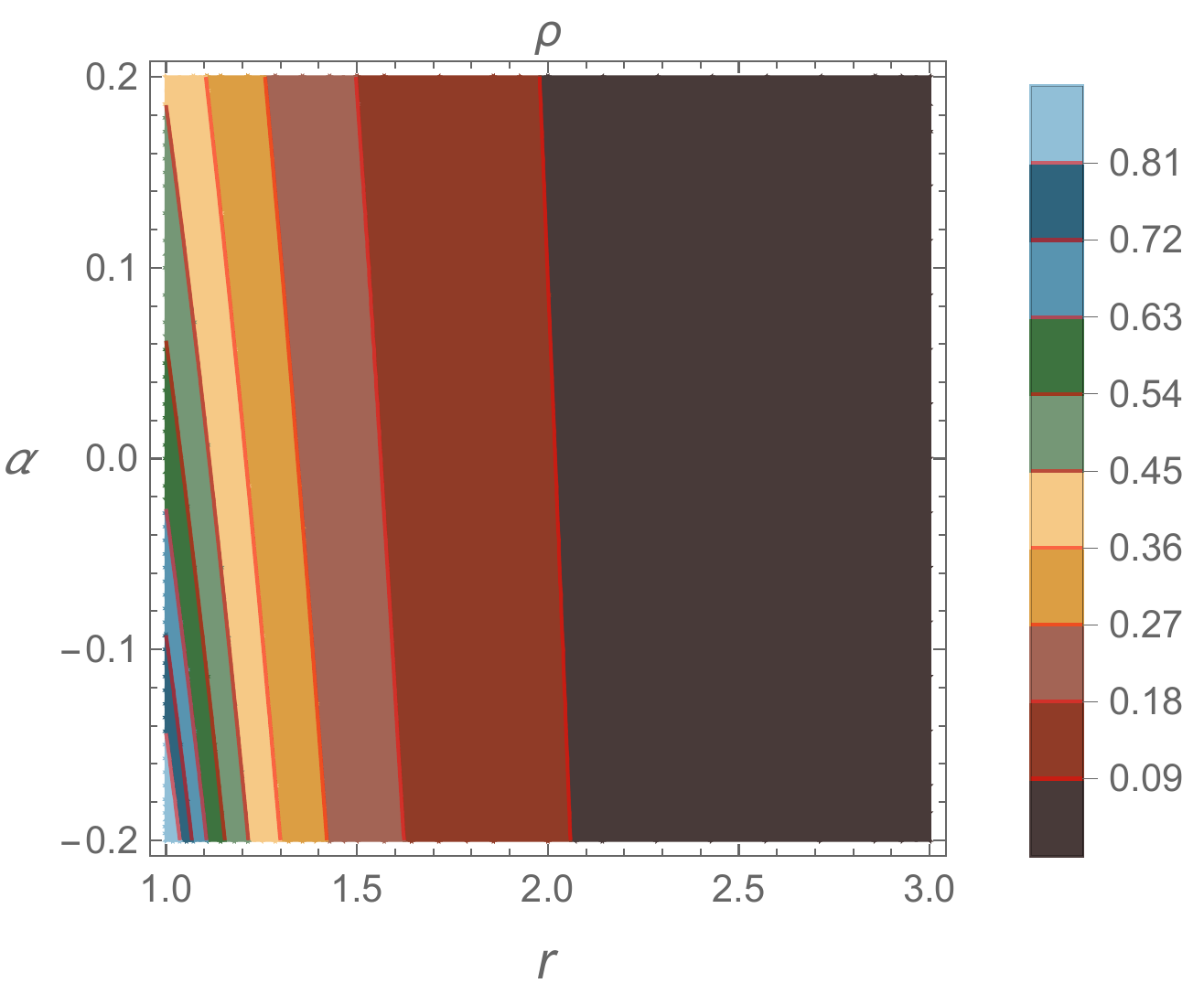}}
	    \subfloat[NEC $\rho+p$\label{fig:Be1}]{\includegraphics[width=0.4\linewidth]{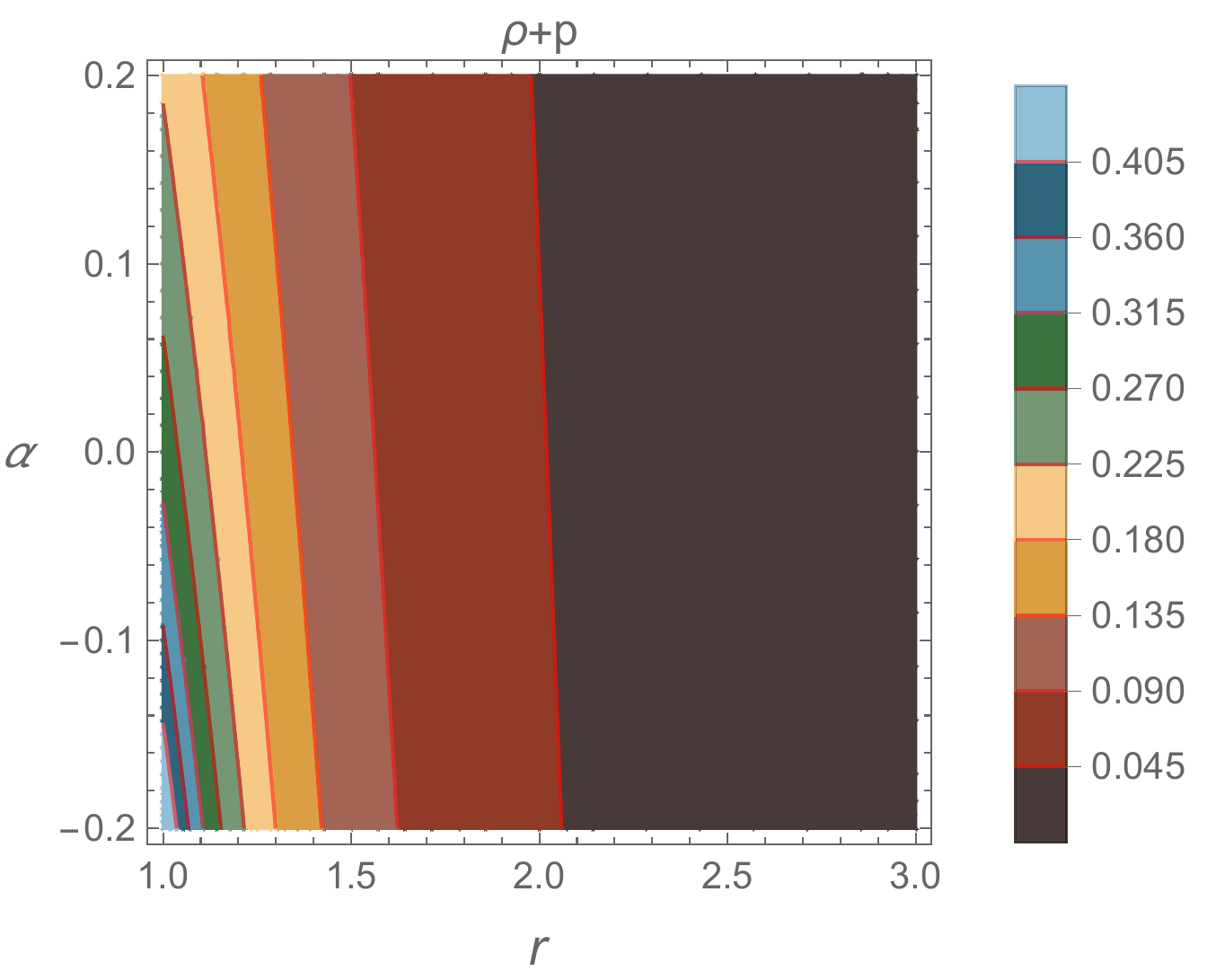}}\\
	    \subfloat[DEC $\rho-|p|$\label{fig:Be2}]{\includegraphics[width=0.4\linewidth]{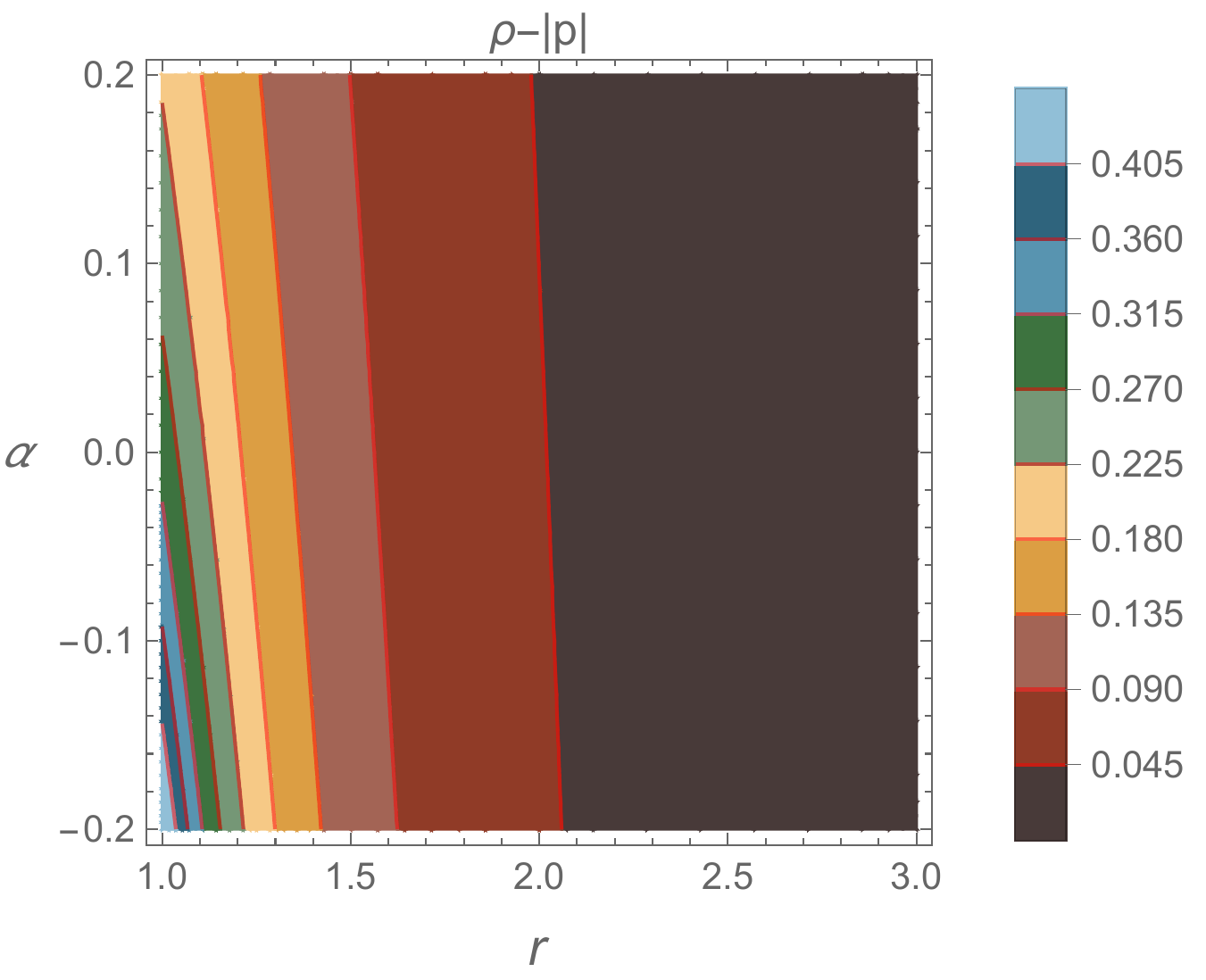}}
	    \subfloat[SEC $\rho+3p$\label{fig:Be3}]{\includegraphics[width=0.4\linewidth]{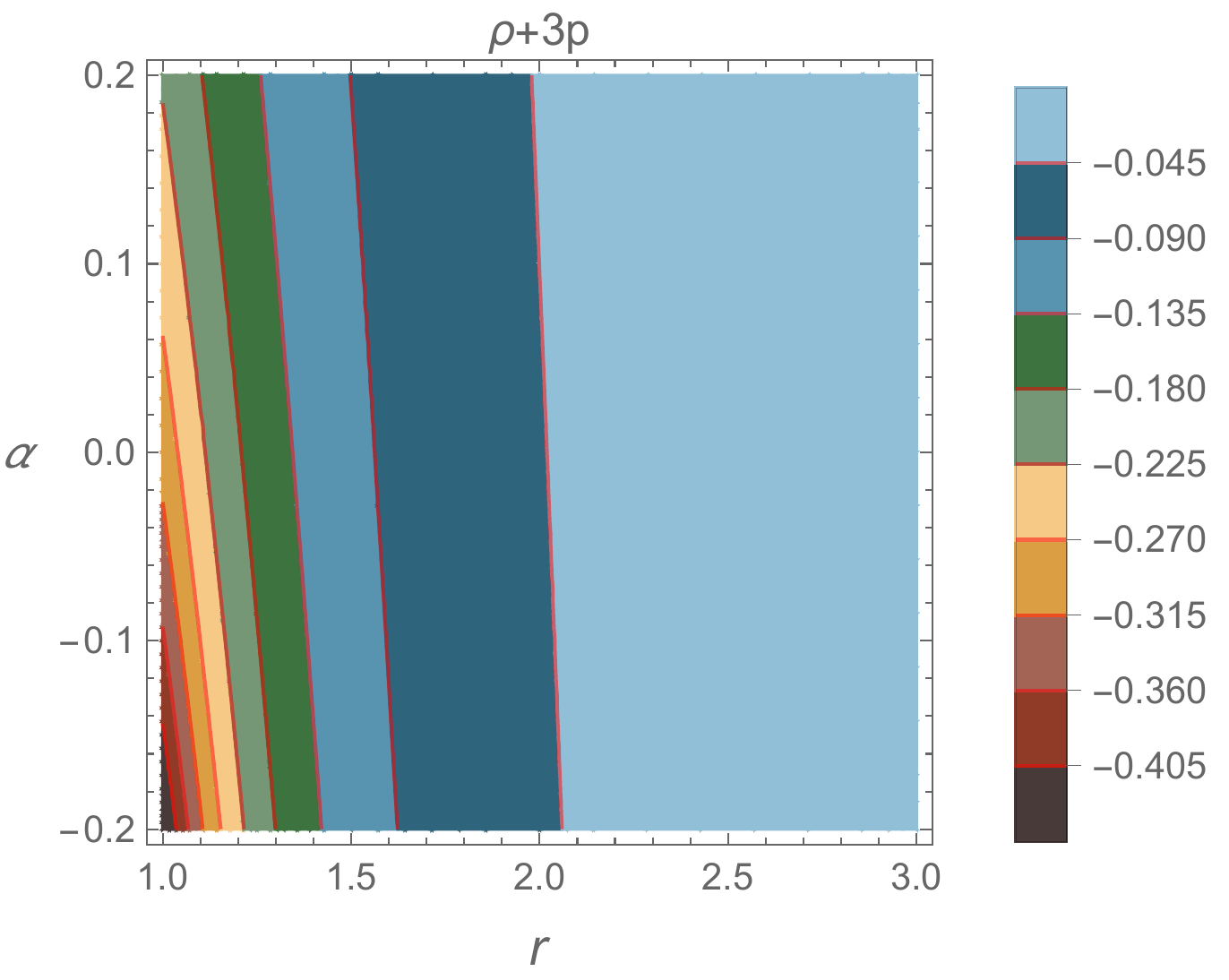}}
	    \caption{Model B: A plot showing the behavior of energy density $(\rho)$, NEC $(\rho+p)$, DEC $(\rho-|p|)$ and SEC $(\rho+3p)$ with $r_0=1$,$m=0.3$ and $\omega=-0.5$.}
	    \label{fig:Bec}
	\end{figure*}
	\begin{table}[h!]
		\caption{Model B: The range of values of the model parameters $\omega$ and $m$  for which $\rho(r)>0, \; \forall\; r$ with $\alpha\in[-0.2,0.2]$.}
		    \label{tab:table1}
		%\begin{ruledtabular}
		    \centering
		    \begin{tabular}{|c|rl|}
		        \hline
		        $m$             & \multicolumn{2}{c|}{$\omega$} \\
		        \hline
		        $(-1,0)$    &  $(-\infty,-1)$ &\textit{(phantom)}\\
		                        &  -1             &\textit{($\Lambda$CDM)}\\
		        \hline
		        $(0,\infty)$        &  $\left(-\frac{2}{3},-\frac{1}{3}\right)$ & \textit{(quintessence)} \\
		                        & 0 &\textit{(dust)}\\
		                        & $\frac{1}{3}$ &\textit{(radiation)}\\
		       \hline
			\end{tabular}
		   %\end{ruledtabular}
		\end{table}
	 \begin{table*}[!]
		\caption{Model B: The interpretation of energy conditions with the set of values for $\omega$, $m$ for which $\rho>0$. Also, the model parameter $\alpha\in[-0.2,0.2]$.}
		    \label{tab:table2}
		%\begin{ruledtabular}
		    \centering
		    \begin{tabular}{|c|c|c|c|c|c|}
		        \hline
		        $m$         & \multicolumn{2}{c|}{$(-1,0)$}    & \multicolumn{3}{c|}{$(0,\infty)$} \\
		        \hline
				$\omega$    & $(-\infty,-1)$    & -1    & $\left(-\frac{2}{3},-\frac{1}{3}\right)$ & 0     &$\frac{1}{3}$\\ 
				\hline
				$\rho+p$    & violated  & satisfied & satisfied & satisfied & satisfied\\
				$\rho-|p|$    & violated  & satisfied  & satisfied & satisfied & satisfied\\
				$\rho+3p$   & violated  & violated  & violated  & satisfied & satisfied\\
				\hline
			\end{tabular}
		   %\end{ruledtabular}
		\end{table*}
      \end{widetext}
    \begin{center}
       \textbf{Wormhole Model B:}
    \end{center}
    \par In the previous section, we determined the shape function by considering the power-law form of energy density. In this section, we shall choose a specific shape function and thus obtain the physical entities such as energy density and pressure. For this case, we assume the isotropic matter distribution $( p_r=p_t=p)$.
    
    \par Here, we take the matter Lagrangian density to be the function of pressure and is defined as $\mathscr{L}_m=p$ \cite{lmp1, lmp2}. This definition of $\mathscr{L}_m$ leads to a significant scenario in $\mathpzc{f}(\mathcal{R},\mathscr{L}_m)$ gravity as it has implications on the non-geodesic motion of the test particles due to the extra force \cite{extraforce}. 
    
    \par Now, for the non-minimal coupling of $\mathpzc{f}(\mathcal{R},\mathscr{L}_m)$ gravity defined in \eqref{mod}, we have to solve field equations \eqref{fe1}-\eqref{fe3}. To this end, to investigate the wormhole model we shall impose yet another condition by taking the linear EoS, $p=\omega \rho, -\infty<\omega<1$. Then from \eqref{fe1}, we get an expression for $\rho$ as,
	\begin{equation}\label{rho}
	    \rho=\frac{b'}{2 \alpha  \omega  b'+4 \alpha  b'+3 r^2 \omega +2 r^2}
	\end{equation}

	\textit{Specific Shape Function:} With the shape function $b(r)=r_0\left(\frac{r}{r_0}\right)^m, -\infty<m<1$ \cite{ec2}, we can find an expression for energy density \eqref{rho}. In \figureautorefname$\;$\ref{fig:Bsf1} one can observe increasing behavior of wormhole shape function for $m=0.25$. Clearly, $b(r)$ obeys the flaring-out condition [\figureautorefname$\;$\ref{fig:Bsf2}], throat condition [\figureautorefname$\;$\ref{fig:Bsf3}], and asymptotic flatness condition [\figureautorefname$\;$\ref{fig:Bsf4}].  Further, the term $1-b(r)/r = 1- r_0 r^{n-1}$ depicting the horizon structure, converges to 1, for a large value of $r$. This quantity is finite for all values in the domain of radial coordinate [\figureautorefname$\;$\ref{fig:Bsf5}].
 
    In this case, we choose the model parameter $\alpha$ in $[-0.2,0.2]$. It is to be noted that, any other value for $\alpha$ can also be considered, but for the present analysis we go with the mentioned range. We restrict our domains of parameter values, $m$ and $\omega$ for which the spacetime has positive density. The \tableautorefname$\;$\ref{tab:table1} gives the set of values for which energy density remains positive.  Further, with the help of \eqref{rho} and linear EoS relation, we can interpret the energy conditions for our wormhole model.  \figureautorefname$\;$\ref{fig:Bec} depicts the profile of energy conditions in quintessence region, particularly for $\omega \in \left(-\frac{2}{3},-\frac{1}{3}\right)$. Interestingly, NEC is satisfied which indicates the absence of exotic fluid. Moreover, by varying the parameter values in the plot we analyzed the behavior of ECs. Readers may refer to \tableautorefname$\;$\ref{tab:table2} that gives the detailed interpretation of different energy conditions for the present wormhole model based on the EoS parameter.

%%%%%%%%%%%%%%%%%%%%%%%%%%%%%%%%%%%%%%%%%%%%%%%%%%%%%%%%%%%%%%%%%%%%%%%%%%%	
	
\section{Final Remarks}\label{VI}

\par A wormhole is an interesting hypothetical spacetime structure whose physical existence is still a big question. Over the past few decades, the research community is thriving in exploring the nature of this mathematically defined geometric element. In the investigation of wormholes, modified theories are found to be more potent than GR. In this regard, the current work assessed tideless traversable wormholes in the background of $\mathpzc{f}(\mathcal{R},\mathscr{L}_m)$ gravity. Here, we have considered a non-minimal coupling  $\mathpzc{f}(\mathcal{R},\mathscr{L}_m)=\dfrac{\mathcal{R}}{2}+(1+\alpha \mathcal{R})\mathscr{L}_m$. Some interesting and viable outcomes of the current study are listed below:
\begin{itemize}
  \item First, with the anisotropic matter distribution and a specific form of energy density, we have derived an expression for shape function. This form of energy density is considered by Kim in \cite{powerlaw1} to validate the flare-out condition for a traversable wormhole.  
  \item It is examined that the obtained shape function satisfies all the necessary and viable conditions [see \tableautorefname$\;$\ref{tab:table0}] for the wormhole existence. The choice for the integral constant $k$ is made on the basis of the throat condition. Thus, $b(r_0)=r_0$ is obeyed. It is interesting to mention that, the derived shape function has satisfied the flaring-out condition under the calculated wormhole throat radius, i.e., $r_{0}=1$.
  \item Further, the involved parameter $\alpha$ is phenomenal in describing the strength of coupling between matter and geometry. In this study, we obtained a condition on the parameter space of  $\alpha$ as $\alpha<0.125$. This implies the existence of weak non-minimal coupling between geometry and matter source.
  \item We analyzed the horizon structure through the interpretation of $1-b(r)/r$ and the summary of these results is given in \tableautorefname$\;$\ref{tab:table0}.
  
  \item Next, for the isotropic case we have considered the linear EoS relation and obtained expressions for physical entities.
\item In the present analysis, we have presumed power-law shape function as $b(r)=r_0\left(\frac{r}{r_0}\right)^m, -\infty<m<1$.
\item  Further, we have verified the energy conditions for both cases. For wormhole model A, NEC was found to be violating, but for model B it is satisfied, while SEC is violated for both models.
\item  The violation of NEC indicates the presence of exotic matter. For ordinary matters, NEC is satisfied. Here, model A confirms the need for exotic matter, while model B shows the presence of non-exotic matter. Recently, Rosa et al., have studied traversable wormhole solutions in the framework of $f(R,T)$ gravity filled with non-exotic matter \cite{res2}. In \cite{res3}, Sadeghi et al., have examined wormhole geometry with Lorenzian distribution in $f(R)$ gravity. In their study, NEC is satisfied.  The interpretation of NEC is significant to deal with exotic matter problems. The obtained result for model B is similar to the study on traversable wormholes in the context of different modified theories \cite{ec1,ec4,ec5, res1}.
\end{itemize}
It is necessary to mention, the present investigated results in the background $\mathpzc{f}(\mathcal{R},\mathscr{L}_m)$ gravity are physically viable. 
\section*{Data Availability Statement}
There are no new data associated with this article.
\begin{acknowledgments}
N.S.K. and V.V. acknowledge DST, New Delhi, India, for its financial support for research facilities under DST-FIST-2019. G. Mustafa is very thankful to Prof. Gao Xianlong from the Department of Physics, Zhejiang Normal University, for his kind support and help during this research. Further, G. Mustafa acknowledges the Grant No. ZC304022919 to support his Postdoctoral Fellowship at Zhejiang Normal University.
\end{acknowledgments}

%\nocite{*}

\end{document}